\documentclass{iopart}
\usepackage{graphics}
\usepackage{xurl}

\begin{document}

\title[Site selection for the Laser Interferometer Lunar Antenna (LILA)]{Site selection for the Laser Interferometer Lunar Antenna (LILA)}

\author{James Trippe$^1$, Ronald Polidan$^2$, Teviet Creighton$^3$, Philippe Lognonn\'e$^{4,5}$, Mark Panning$^6$, Volker Quetschke$^3$, Kris Izquierdo$^7$, Brett Shapiro$^7$, and Karan Jani$^1$}
\address{$^1$ Vanderbilt University}
\address{$^2$ Lunar Resources, Inc.}
\address{$^3$ University of Texas Rio Grande Valley}
\address{$^4$ Université Paris Cité, Institut de physique du globe de Paris, CNRS}
\address{$^5$ California Institute of Technology}
\address{$^6$ National Aeronautics and Space Administration}
\address{$^7$ Johns Hopkins Applied Physics Laboratory}
\ead{james.m.trippe@vanderbilt.edu}

\date{\today}

\begin{abstract}
  The Earth's Moon presents a uniquely advantageous environment for detecting astrophysical gravitational waves (GWs), particularly in the decihertz regime. The Laser Interferometer Lunar Antenna (LILA) project plans to perform GW measurements on the lunar surface, using the Moon's seismic quietness to access this band. Two mission concepts are considered: the proof-of-concept 1 km initial LILA (iLILA) and the full LILA Observatory, whose equilateral arms are at least 40 km long. The Moon's changing orientation and orbital motion provide the time-dependent source modulation needed for sky localization, so the detector response does not impose a unique lunar region. Practical considerations, most critically line of sight (LOS), nevertheless constrain deployment. A coarse-to-fine search of existing lunar datasets identifies and ranks candidate grid locations. Of 4,050 global grid seeds, 434 iLILA seeds and 71 Observatory circumcenters pass both search stages. These results demonstrate that sites suitable for GW detection naturally exist in large numbers on the lunar surface.
\end{abstract}


\section{Introduction}
\label{s:introduction}

The Moon is well suited to gravitational-wave (GW) detection ~\cite{2020arXiv200708550J,Harms:2021}. LILA comprises the Lunar Deformation Channel (LDC) ~\cite{Creighton2025} and Geodesic Deformation Channel (GDC) ~\cite{shapiro2025vibrationisolationlaserinterferometer}, which exploit the quiet lunar seismic environment for astrophysical measurements and lunar normal-mode science ~\cite{panning2025potentiallunarinteriorscience}. The program matures these concepts through three missions: the payload-independent LILA Autonomous Interferometer Technology Demonstration; the 1 km LDC implementation iLILA, which targets lunar normal modes and seismic-noise risk reduction; and the LILA Observatory, combining LDC and GDC for planetary and astrophysical science. Only the latter two impose site-selection requirements and are considered here.

The key difference between the concepts in the context of site selection is their arm geometry: iLILA is a pair of 1 km interferometric strainmeters, while LILA Observatory requires an equilateral triangular interferometer with arms at least 40 km long. The primary detection band of interest for LILA is in the decihertz regime, between the millihertz regime of the Laser Interferometer Space Antenna (LISA) ~\cite{LISAmain} and the hertz regime of terrestrial detectors such as the Laser Interferometer Gravitational-Wave Observatory (LIGO) ~\cite{LIGO:2012aa,LIGO-Instrument2018,Aasi2015AdvancedLIGO}. Sky localization in the decihertz band can use the Moon's time-dependent orientation and motion ~\cite{2020arXiv200708550J}; consequently, no unique lunar region is required by the detector response itself. Site selection is instead governed by deployment, environmental, and engineering constraints.

The primary constraint for both concepts is LOS between stations. Other considerations include lunar terrain vehicle (LTV) limits, shallow moonquakes, micrometeoroid seismic hum, anthropogenic noise, and environmental protection. While their context to the LILA missions are newly studied here, all of these constraints are well known, as indicated with references provided in each subsection of Section ~\ref{s:constraints}. This section also translates these considerations into the hard constraints and soft preferences used by the search.

\section{Considerations for Site Selection}
\label{s:constraints}

\subsection{Line of Sight}

Both iLILA and LILA Observatory require unobstructed LOS between optical stations. The search accounts explicitly for the Moon's curvature and sampled topography rather than applying a flat-terrain horizon approximation. Let $\Gamma$ be the central angle between two stations and $\alpha$ the central angle from the first station to a terrain-profile sample. The endpoint radii are $r_0=R_{\rm Moon}+z_0+h$ and $r_1=R_{\rm Moon}+z_1+h$, where $z_0$ and $z_1$ are local terrain elevations, $R_{\rm Moon}=1737.4$ km, and each station is modeled at $h=1$ m above its local ground. The radial position and altitude of the straight optical chord are
\begin{equation}
r_{\rm LOS}(\alpha)=
\frac{r_0r_1\sin\Gamma}
{r_1\sin(\Gamma-\alpha)+r_0\sin\alpha},
\qquad
z_{\rm LOS}(\alpha)=r_{\rm LOS}(\alpha)-R_{\rm Moon}.
\label{eq:spherical_los}
\end{equation}
At each interior profile sample, the clearance is $c(\alpha)=z_{\rm LOS}(\alpha)-z(\alpha)$. The minimum sampled interior clearance must be at least 1 m on both iLILA arms and all three Observatory sides.

iLILA additionally requires an absolute terrain-elevation difference between the center and each endpoint of
\begin{equation}
|\Delta z|\geq L\tan(5^\circ).
\label{eq:ilila_height_floor}
\end{equation}
For the fixed $L=1$ km arms, the effective threshold is 87.49 m. This endpoint-height requirement is separate from the 1 m interior LOS clearance. LILA Observatory has no analogous endpoint-height-difference requirement or additional angular margin; feasibility requires only the three spherical LOS tests for stations 1 m above their respective local terrain. iLILA uses additional margin to identify a robust field of co-manifestable sites, whereas LILA Observatory applies unbuffered feasibility thresholds because its boutique site will be selected and finalized through mission-specific evaluation.

\subsection{Capabilities of Deployment Tools}
iLILA may be deployed using a lunar terrain vehicle (LTV) ~\cite{NASA_ltv}, which provides a representative mobility model for next-generation deployment. The search imposes a 10 km total route limit for two center-to-endpoint out-and-back deployments and a $15^{\circ}$ maximum local traversed slope. Each one-way route may detour within a 250 m half-width corridor around the corresponding straight optical arm; the optical arm itself does not detour. These are study assumptions rather than published vehicle performance guarantees. No LTV constraint is applied to LILA Observatory, whose deployment architecture is outside the scope of this search.

\subsection{Shallow Moonquakes}

Shallow moonquakes may reduce duty cycle and obscure normal-mode signals, unlike scientifically useful deeper events ~\cite{LOGNONNE201565}. Their spatial coupling to LILA cannot yet be modeled credibly. Because shallow seismicity has been associated with lobate scarps ~\cite{WATTERS2010}, the search uses mapped scarp distance as a conservative proxy: 50 km is a hard keepout and greater distance is preferred.

\subsection{Micrometeoroids}

Micrometeoroid impacts generate a lunar seismic hum ~\cite{Lognonne2009} that may affect the decihertz band, although common-mode rejection should suppress it for the low frequency LDC concept \cite{Harms:2022}. For the high frequency GDC, seismic isolation such as found in Advanced LIGO should be sufficient \cite{Aasi2015AdvancedLIGO}. Current models do not support site-specific coupling. Because flux peaks near the lunar equator \cite{LEFEUVRE20111}, the latitude preference favors mid-high absolute latitudes.

\subsection{Geology}
Regolith thickness varies across the Moon ~\cite{https://doi.org/10.1029/2001JE001506}, and both LDC coupling and GDC isolation require a stable foundation. Available thickness estimates cover only limited locations ~\cite{RajsicEtAl2026_Regolith}. An attempt was made to interpolate regolith thickness between sites, but cross-method holdout did not preserve the measured ordering. Thus, as there is no credible lunar-wide dataset, regolith thickness was ignored in this search.

\subsection{Anthropogenic Noise}
Terrestrial experience shows that human activity can contaminate interferometer data; logging near LIGO Livingston, for example, produced noise mainly at 1--6 Hz \cite{Glanzer_2023}. Lunar operations could also introduce lower-frequency disturbances. The poles are major exploration targets ~\cite{NASA_SP_20205009602_2020}, while lava tubes ~\cite{Horvath2022}, pyroclastic deposits ~\cite{Milliken2017}, and titanium-rich mare ~\cite{SATO2017216} may attract nonpolar activity. Because future activity and mitigation schedules are uncertain, anthropogenic noise is not scored.

\subsection{Protection from the Lunar Environment}
The lunar environment poses risks from dust ~\cite{Rubanu_2009,NASA2023}, radiation ~\cite{ARAUJO2005451,DEANGELIS2007169}, temperature ~\cite{Horanyi2014,Horanyi:2015}, and magnetic-field anomalies ~\cite{Tsunakawa2010}. Latitude-dependent environmental exposure is treated as a soft preference because engineering mitigation is possible. Mapped magnetic anomalies are instead subject to a hard keepout, as described below.

\subsubsection{Dust}
Dust is not expected to limit detector sensitivity ~\cite{Creighton2025}, but can obscure optics and affect actuators ~\cite{Katzan1991}. Electrostatic lofting may increase near the moving terminator ~\cite{Horanyi:2015}, so operational mitigation may be required regardless of site.

\subsubsection{Radiation}
Radiation can charge test masses ~\cite{ARAUJO2005451} and degrade optics ~\cite{Johnston2003}. Lunar exposure includes solar particles, albedo, and galactic cosmic rays ~\cite{DEANGELIS2007169}; polar illumination geometry and permanently shadowed regions (PSRs) can reduce the solar component ~\cite{GLASER201478,REITZ201278}. This consideration also supports a near-polar preference.

\subsubsection{Temperature}
Surface temperatures range from below 10 K to nearly 400 K ~\cite{WILLIAMS2017300}; even PSRs can vary by approximately 30 K ~\cite{Williams2019}, while temperatures around 30 cm below the surface are much more stable ~\cite{https://doi.org/10.1029/2020JE006623}. Space-qualified optics and electronics have finite temperature ranges ~\cite{EEE-INST-002,MIL-PRF-38534}, and interferometers drift thermally ~\cite{Aasi2015AdvancedLIGO}. Thermal control, subsurface coupling, or duty cycling can mitigate these effects ~\cite{NASA_PTC_Guidebook_4_0_2023}. The latitude score therefore avoids both the equatorial thermal extreme and the pole, where PSRs are very cold and moving shadows can drive local variation ~\cite{MAZARICO20111066}.

\subsubsection{Magnetic fields}
The lunar crustal magnetic field is spatially heterogeneous ~\cite{Tsunakawa2010}. In the search product, cells with field magnitude at or above 5 nT in the 30 km-altitude map define the mapped anomaly set, and every covered physical station must lie at least 50 km from that set. Locations without reliable map coverage remain eligible but are flagged; magnetic distance has zero ranking weight and therefore cannot penalize an unmapped location.

\subsection{Other considerations}
A few final considerations are not modeled in the present iLILA or LILA Observatory search but are noted for completeness.

\begin{itemize}
    \item Communications are currently easier on the lunar near side, while relay architectures are being developed for far-side operations ~\cite{apl_lsii_2025_lit}. The search nevertheless leaves both hemispheres eligible; a mission-specific communications constraint can be applied later.
    \item The Moon's surface deforms through tides and thermal cycling, producing low-frequency strainmeter signals ~\cite{https://doi.org/10.1002/2013JE004559}. Typical tidal vertical displacement is 10--30 cm \cite{Thor2020}, with monthly strain of order $6\times10^{-8}$ \cite{Pou2024}. The response depends on orbital eccentricity, inclination, and lunar libration and therefore is not canceled at the poles. Adequate dynamic range and model-based subtraction will be required, as for planetary very-broadband seismometers \cite{Lognonn2019,Lognonn2020}. \item Static gravity-field differences measured by GRAIL ~\cite{Lemoine2014GRGM900C,Wieczorek2013GRAILCrust,GRAILDerivedGravityPDS} do not enter the common-mode GW measurement. These are ignored.
    \item Mission-specific constraints such as co-manifestation, power availability, and crew access are deferred until an architecture is selected.
\end{itemize}

\subsection{Site Location Constraints} 
\label{ss:constraints}
The search treats~\textbf{required} conditions as pass/fail constraints and~\textbf{preferred} conditions as ranking terms that can be mitigated by engineering design.
\vspace{0.5 cm}

\begin{center} \underline{iLILA}
\end{center} 
\begin{itemize}
    \item An L-shaped pair of 1 km arms separated by $90^{\circ}$ is~\textbf{required}.
    \item A minimum 87.49 m absolute terrain-elevation difference between the center and each arm endpoint is~\textbf{required}.
    \item Stations 1 m above local terrain and at least 1 m sampled interior LOS clearance on each arm are~\textbf{required}.
    \item Two center-to-endpoint out-and-back LTV routes totaling at most 10 km, with traversed slope $\leq15^{\circ}$ and detours confined to a 250 m arm corridor, are~\textbf{required}.
    \item A graded latitude preference that peaks at $|\phi|=75^{\circ}$ and falls to zero at both the equator and poles is~\textbf{preferred}.
    \item A 50 km mapped magnetic-anomaly keepout is~\textbf{required}.
    \item A 50 km mapped lobate-scarp keepout is~\textbf{required}; greater mapped distance up to 200 km is~\textbf{preferred}.
    \item Lower optical-profile roughness is~\textbf{preferred}.
\end{itemize}
\vspace{0.5 cm}
\begin{center} \underline{LILA Observatory}
\end{center} 
\begin{itemize}
    \item A spherical equilateral triangle with side lengths from 40 km to approximately 69.3 km and fractional side-length spread $\leq0.1\%$ is~\textbf{required}.
    \item Stations 1 m above local terrain and at least 1 m sampled interior LOS clearance on all three sides are~\textbf{required}.
    \item The same graded latitude preference used for iLILA is~\textbf{preferred}.
    \item A 50 km mapped magnetic-anomaly keepout is~\textbf{required}.
    \item A 50 km mapped lobate-scarp keepout is~\textbf{required}.
\end{itemize}

The search algorithm used to apply these requirements and perform the scoring is detailed in the next section.

\section{Search Methodology}
\label{s:search}

The search has different objectives for the two LILA concepts. For iLILA, it tests whether the registered engineering assumptions leave deployment options across a wide range of lunar regions. For LILA Observatory, it tests whether any three-station configurations satisfy the more restrictive long-baseline LOS requirement. The output is a reconnaissance catalog: a modeled 1 m clearance on the available rasters is not a survey-grade clearance guarantee, and any selected location requires mission-specific topographic and geotechnical follow-up.

\subsection{Lunar data sources}
\label{ss:lds}

The production search uses the following lunar products:

\begin{itemize}
\item \textbf{LRO LOLA global LDEM118}: the 256 pixels-per-degree (approximately 118 m posting at the equator) elevation product used for the global screen. The raster is read in native-resolution windows without terrain downsampling; elevation profiles derived from it support geometry, height, LOS, slope, route, and roughness calculations ~\cite{USGS_LOLA_LDEM118,NASAPDS_LOLA_GDR}.
\item \textbf{SLDEM2015 and adjusted polar LOLA DTMs}: approximately 60 m regional products used for fine inspection at nonpolar and polar locations, respectively ~\cite{MIT_SLDEM2015_Data,BarkerEtAl2016_SLDEM2015,NASAPGDA_LOLAPoles,BarkerEtAl2025_LOLAPolarRoughness,BarkerEtAl2021_ImprovedLOLA}.
\item \textbf{LROC lobate-scarp catalogs}: nonpolar and polar catalogs used to apply the mapped 50 km scarp keepout and the scarp-distance preference ~\cite{LROC_LobateScarps,NelsonEtAl2014_LobateScarps,LROC_PolarScarpLocations,WattersEtAl2015_LunarScarps}.
\item \textbf{NASA PDS lunar crustal magnetic-field map}: the 30 km-altitude field product used to define mapped anomalies and apply the 50 km magnetic keepout ~\cite{HoodEtAl2020_MagneticMapData,HoodEtAl2021_MagneticMap}.
\item \textbf{Regolith-thickness estimates}: the Raj\v{s}i\'{c} data are retained as a descriptive field and for validation experiments, but carry zero production weight and do not affect acceptance or ranking ~\cite{RajsicEtAl2026_Regolith,JonesRajsic2026_RegolithData}.
\end{itemize}

Scarp and magnetic products have incomplete geographic coverage. An uncovered location is retained and flagged rather than treated as either a measured pass or failure. Its associated distance score is set to zero. The magnetic-distance ranking weight is also zero for covered locations, so the magnetic product acts only as a mapped keepout.

\subsection{Coarse global search}

Both concepts use the same phase-aligned lattice of 4,050 global grid seeds. Neighboring seed indices are separated by 1,024 LDEM pixels, equivalent to $4^{\circ}$ in latitude and longitude and approximately 121.29 km at the equator; east--west physical separation decreases with $\cos\phi$. Terrain within each candidate geometry is nevertheless sampled from native LDEM118 windows, with LOS profiles sampled no more sparsely than 100 m. Configuration counts are internal search trials and are not treated as independent statistical observations. The results are considered existence evidence on tested grid cells for a valid configuration; as this initial search grid is larger than the arm lengths, there are likely several false negatives introduced by this coarse stage search. 

At each iLILA seed, the code tests 36 orientations from $0^{\circ}$ to $350^{\circ}$ in $10^{\circ}$ increments. Each orientation has two 1 km arms separated by $90^{\circ}$. Both arms must satisfy the 87.49 m effective endpoint-height threshold, 1 m spherical LOS clearance, and 15$^{\circ}$ route-slope limit. An LTV pathfinder may detour within 250 m of the straight optical arm when the direct route is too steep. Twice the sum of the two one-way route lengths must not exceed 10 km. The scarp and magnetic keepouts are evaluated at the physical iLILA center.

For LILA Observatory, the common grid seed is only the computational circumcenter of a trial triangle; it is not a physical station, and its elevation is neither sampled nor scored. The code tests 12 unique rotations at each of seven side lengths: 40--65 km in 5 km increments plus the exact upper length, approximately 69.3 km, implied by the 80 km spherical-circumcircle diameter. This gives 84 configurations per grid seed. The three sides must each be at least 40 km, with a 10 m numerical tolerance applied only to that lower-bound comparison, and their fractional length spread must not exceed 0.1\%. Each physical station is modeled 1 m above local terrain, and all three spherical optical chords must clear sampled interior terrain by at least 1 m. Environmental keepouts are evaluated at all three stations. No station-height-difference, LTV, slope, or roughness requirement is imposed on the Observatory.

\begin{figure}
\begin{center}
\resizebox{400pt}{!}{\includegraphics{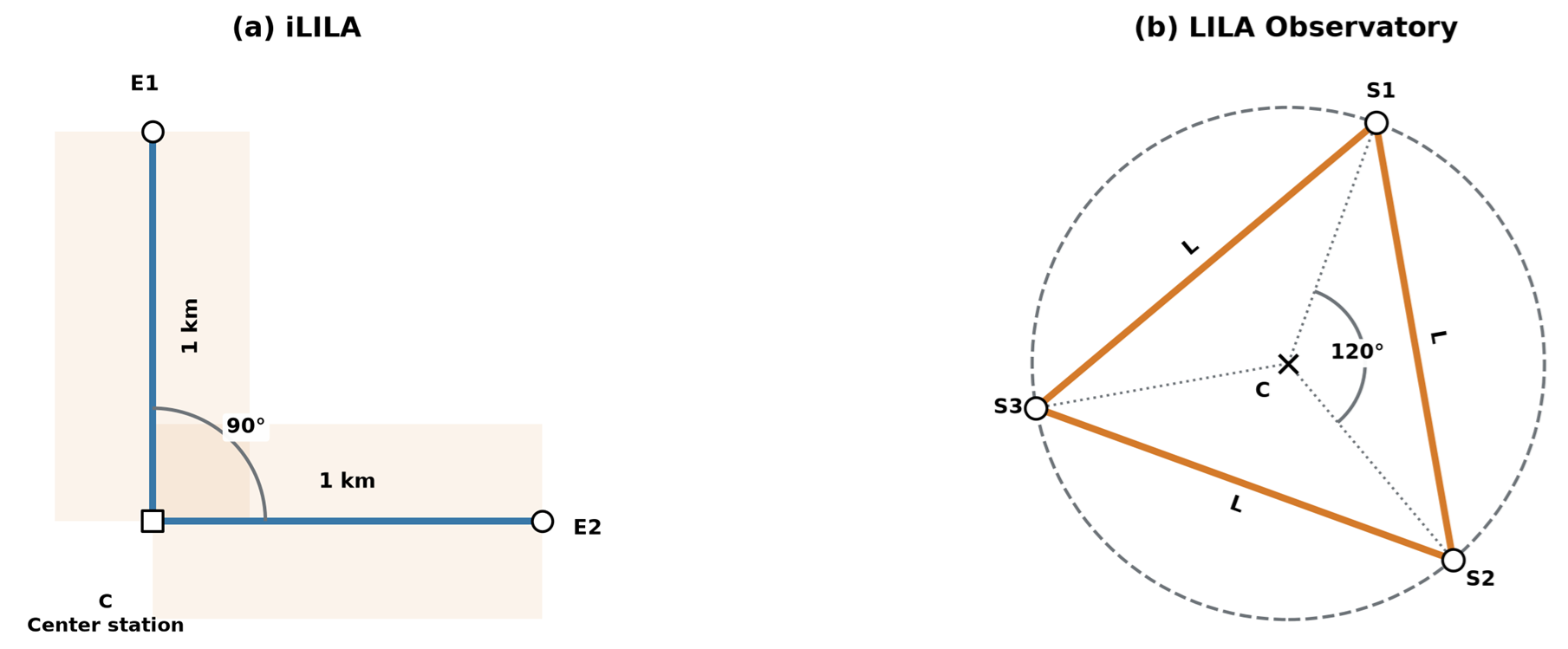}}
\end{center}
\caption{\label{fig:Geometry}
  Conceptual search geometries. iLILA uses two fixed 1 km arms separated by $90^{\circ}$, with a 250 m LTV route corridor around each straight optical arm. LILA Observatory uses a spherical equilateral triangle with side lengths from 40 km to approximately 69.3 km; the plotted center is a computational circumcenter, not a physical station.}
\end{figure}

\subsection{Regional fine inspection}

Every coarse-accepted grid seed advances to a declared regional DTM. iLILA performs a new local center-and-orientation search within 1 km of each coarse seed. Center spacing is nominally 60 m. A hierarchical angular search begins at $10^{\circ}$ and refines to $2^{\circ}$, retaining bounded sets of the strongest intermediate configurations. A final configuration must survive cardinal 60 m center perturbations and $\pm2^{\circ}$ orientation perturbations. The fine search exits after finding a robust passing configuration for that grid seed. The 1 km arm length, $90^{\circ}$ opening angle, 250 m route corridor, and all coarse acceptance thresholds remain unchanged.

For the Observatory, every coarse-feasible triangle is frozen and all three LOS paths are re-evaluated on the corresponding regional DTM. All coarse-feasible triangles are inspected; a grid seed is finally accepted if at least one of its triangles passes the fine terrain test. The Observatory likewise uses the same 1 m station height and 1 m LOS threshold at both stages, with no extra coarse or fine height margin.

\subsection{Ranking}

Hard constraints are applied before scoring and cannot be offset by a high preference score. For each accepted configuration, the shared site score is
\begin{equation}
S_{\mathrm{site}}=1.0L+0.3F,
\label{eq:site_preference_score}
\end{equation}
where $L$ is the normalized latitude preference and $F$ is the normalized mapped scarp-distance preference.

For latitude $\phi$,
\begin{equation}
L(\phi)=
\left\{
\begin{array}{ll}
|\phi|/75^{\circ}, & 0^{\circ}\leq|\phi|\leq75^{\circ},\\
(90^{\circ}-|\phi|)/15^{\circ}, & 75^{\circ}<|\phi|\leq90^{\circ}.
\end{array}
\right.
\label{eq:latitude_score}
\end{equation}
Thus $L=0$ at the equator and poles and $L=1$ at $|\phi|=75^{\circ}$. If $d_{\mathrm{scarp}}$ is the great-circle distance to the nearest mapped lobate scarp, then
\begin{equation}
F={\rm clip}_{[0,1]}
\left(\frac{d_{\mathrm{scarp}}-50~\mathrm{km}}{150~\mathrm{km}}\right).
\label{eq:scarp_score}
\end{equation}
A covered site inside 50 km is rejected before ranking, while a covered site reaches $F=1$ at 200 km. An uncovered site has $F=0$. For the Observatory, the least favorable station value is used for each shared preference so one favorable station cannot mask another.

The iLILA total score is
\begin{equation}
S_{\mathrm{iLILA}}=S_{\mathrm{site}}
+(0.1~\mathrm{m}^{-1}) (C_{\min} - Q)
+(0.01~\mathrm{deg}^{-1})\theta_{\mathrm{margin}},
\label{eq:ilila_score}
\end{equation}
where $C_{\min}$ is the minimum interior LOS clearance across the two optical arms. The mean optical-profile roughness $Q$ is calculated by subtracting a 100 m moving-average profile from each arm, taking the population standard deviation of each residual profile, and averaging the two results. For route segment $j$ on arm $a$, the local slope is
\begin{equation}
\theta_{a,j}=\tan^{-1}\left(\frac{|\Delta z_{a,j}|}{\Delta s_{a,j}}\right),
\end{equation}
where $\Delta s_{a,j}$ is the surface distance between adjacent route samples. If $\theta_{\max,a}$ is the largest traversed slope on route $a$, then $\theta_{\mathrm{margin}}=\min_a(15^{\circ}-\theta_{\max,a})$. 

The Observatory score is
\begin{equation}
S_{\mathrm{Obs}}=S_{\mathrm{site}}
+(1.25\times10^{-4}~\mathrm{m}^{-1})C_{\min},
\label{eq:observatory_score}
\end{equation}
where $C_{\min}$ is the minimum interior clearance over all three triangle sides. Observatory relief and roughness are not score terms. The small clearance coefficient acts only as a tie-breaker after all three LOS paths have passed.

The weights encode engineering judgment rather than an optimized physical loss function. Latitude is the reference-scale shared preference because it represents several environmental considerations; scarp distance is secondary. For iLILA, clearance is the principal continuous engineering margin, while route-slope margin and roughness provide smaller mobility and terrain preferences. For the Observatory, ranking is governed primarily by latitude and mapped scarp distance, with clearance providing only a modest secondary contribution. Regolith and magnetic distance have zero production weight for the data-quality reasons stated above.

\section{Search Results}
\label{s:results}

\subsection{Candidate locations and attrition}

The search identifies feasible screening candidates rather than construction-ready sites. Both concepts were evaluated on the same 4,050-seed global lattice. For iLILA, 145,800 coarse configurations were tested, of which 2,839 configurations at 722 unique seeds were feasible. Fine regional search inspected all 722 seeds, rejecting 288 and accepting 434. For LILA Observatory, 340,200 coarse triangles were tested; 370 triangles at 114 unique circumcenters were coarse-feasible. Fine inspection re-evaluated all 370 triangles, accepting 179 triangles at 71 circumcenters and rejecting the remaining 43 circumcenters. The final lattice yields are therefore 10.72\% for iLILA and 1.75\% for LILA Observatory; conditional on coarse acceptance, the fine-stage yields are 60.11\% and 62.28\%, respectively.

The accepted locations are geographically nonuniform. Both concepts avoid many broad, flat mare and crater-floor regions because the required endpoint relief or long-baseline LOS is unavailable there. iLILA candidates remain distributed over both hemispheres wherever localized relief coexists with traversable deployment routes. Observatory candidates are substantially sparser and favor terrain that leaves three long optical chords simultaneously unobstructed. The highest-ranked sites for both concepts tend toward mid-high absolute latitudes, especially in the southern hemisphere, because the shared latitude preference peaks at $|\phi|=75^{\circ}$.

Several limitations qualify these yields. The terrain tests inherit the posting, interpolation, vertical uncertainty, and coverage of the source DTMs; the nominal 1 m LOS threshold is a model criterion rather than a field guarantee, and sub-grid obstacles or slopes require higher-resolution survey. The finite seed lattice and bounded center, orientation, and arm-length searches are not exhaustive. Scarp and magnetic constraints are enforced where coverage exists, while unmapped locations remain eligible and flagged. Finally, the coefficients encode explicit engineering assumptions rather than a calibrated physical utility, so the catalog is a traceable down-selection product rather than proof of site readiness. The approximately 69.3 km Observatory upper arm length also excludes potentially feasible larger triangles.

\begin{figure}
\centering
\resizebox{1\linewidth}{!}{
\includegraphics{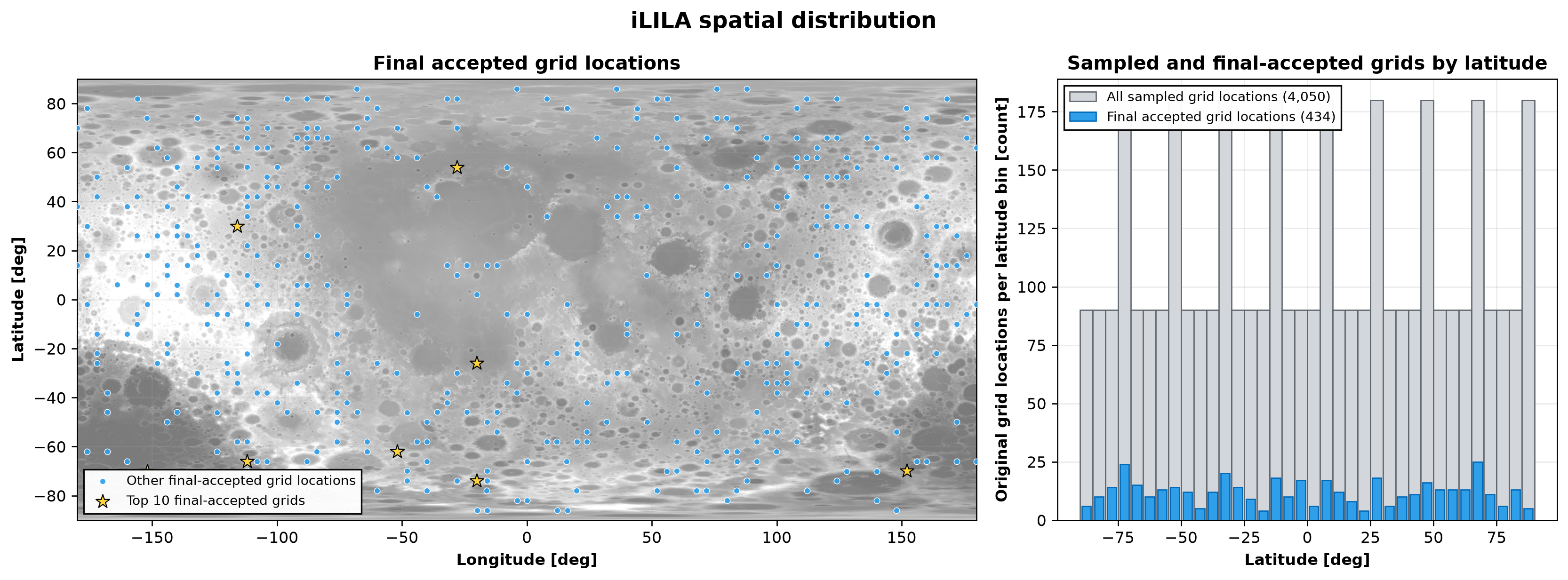}
}
\\[1ex]
\resizebox{1\linewidth}{!}{
\includegraphics{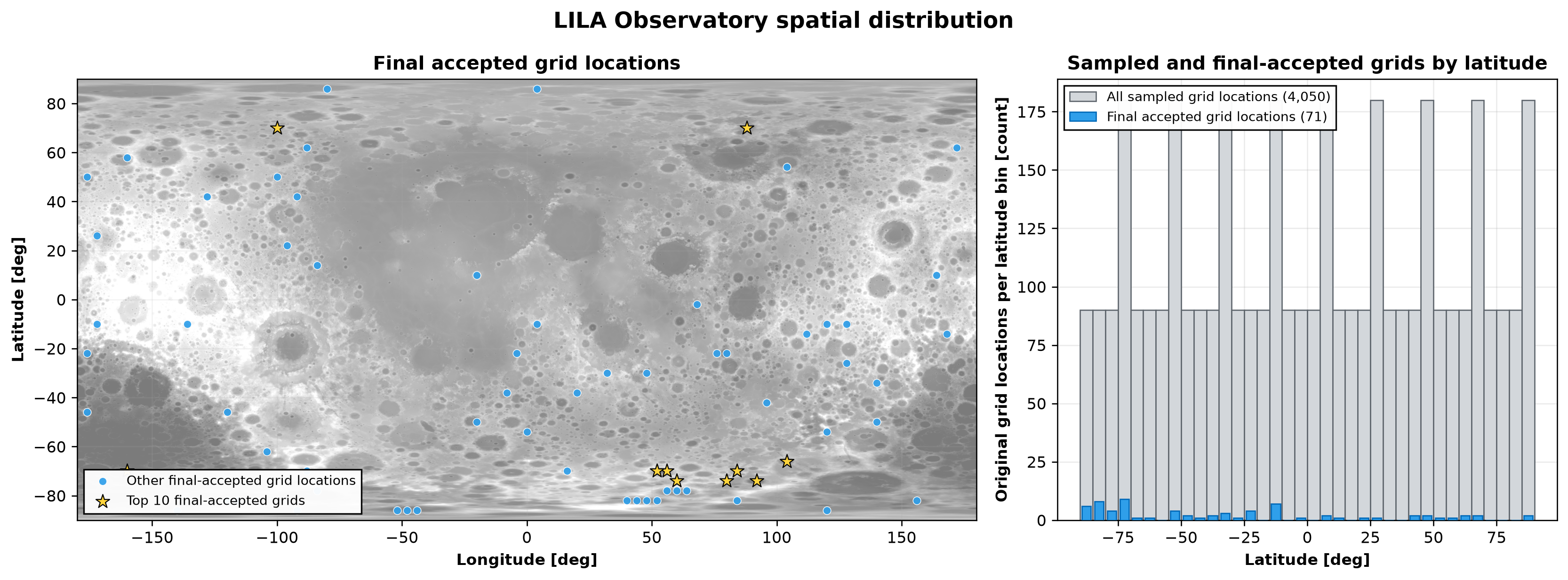}
}
\caption{\label{fig:Distributions}
  Final fine-accepted iLILA (top) and LILA Observatory (bottom) grid locations (blue circles), with the ten highest-ranked grid locations marked by yellow stars. The right panels compare final accepted and total sampled seeds by latitude.
  }
\end{figure}

\subsection{Constraint selectivity}

Hard-constraint selectivity is reported using one record per original global seed, not one record per tested configuration. A coarse physical criterion blocks a seed only when none of the internally tested configurations passes that criterion; several criteria may block the same seed, so rejection percentages overlap and do not sum to the total attrition. Observatory fine failures are assigned to the re-evaluated three-side LOS criterion. The iLILA fine ledger does not uniquely assign every rejected regional search to one physical criterion, so those 288 rejections are shown in pipeline attrition rather than forced into a misleading selectivity category.

For iLILA, endpoint height difference blocks 52.7\% of the global seeds and LOS blocks 39.7\%; mapped scarp and magnetic keepouts block 4.0\% and 1.9\%, respectively. LTV route slope and route length each block only 0.3\%, with perfect overlap. For the Observatory, three-side LOS blocks 97.9\% after coarse and fine evidence are combined, while scarp and magnetic keepouts block 5.0\% and 2.2\%. Thus, terrain geometry rather than mapped environmental keepouts is the dominant hard selector for both concepts.

\begin{figure}
\centering
\resizebox{0.7\linewidth}{!}{
\includegraphics{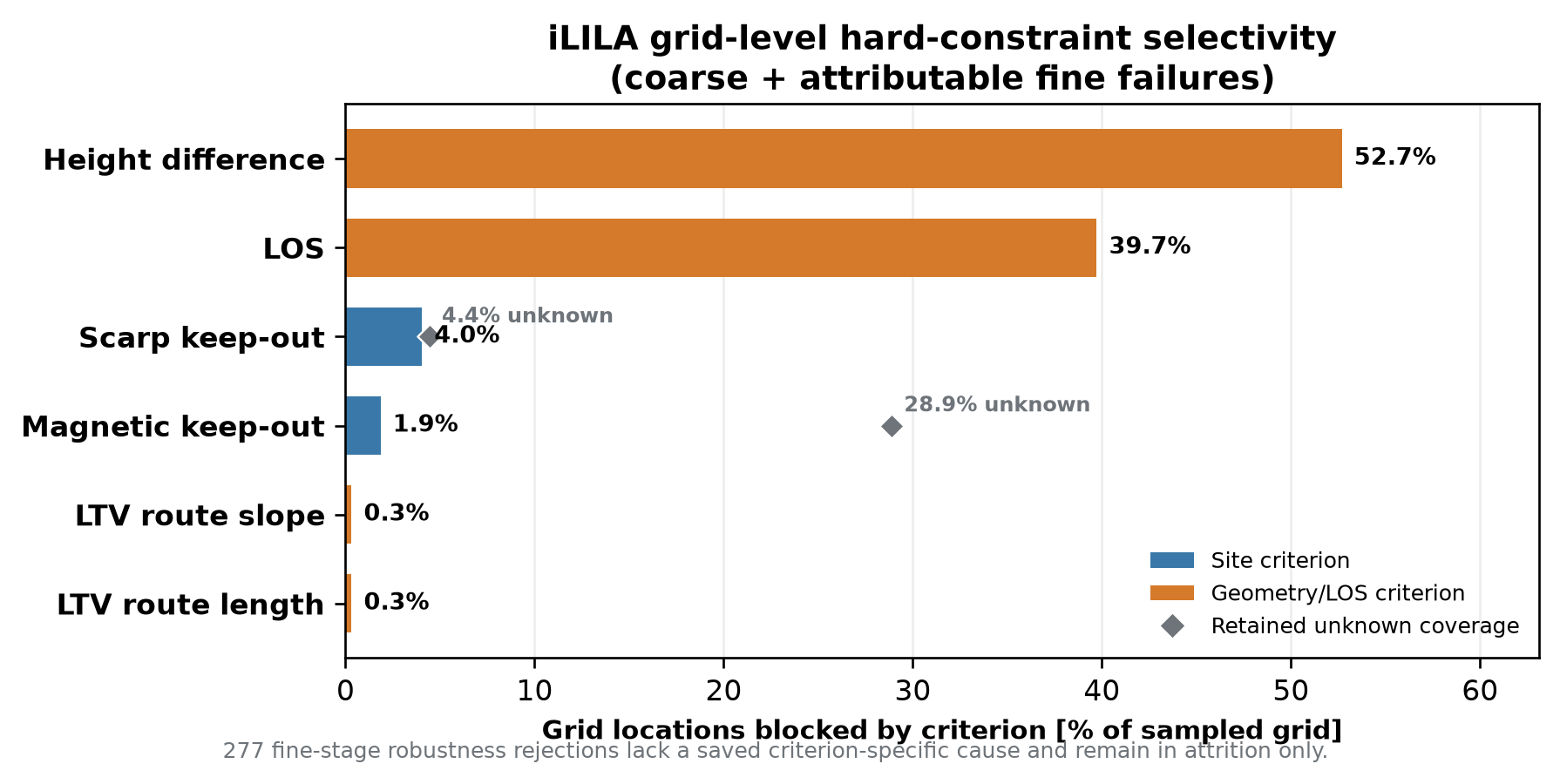}
}
\\[1ex]
\resizebox{0.7\linewidth}{!}{
\includegraphics{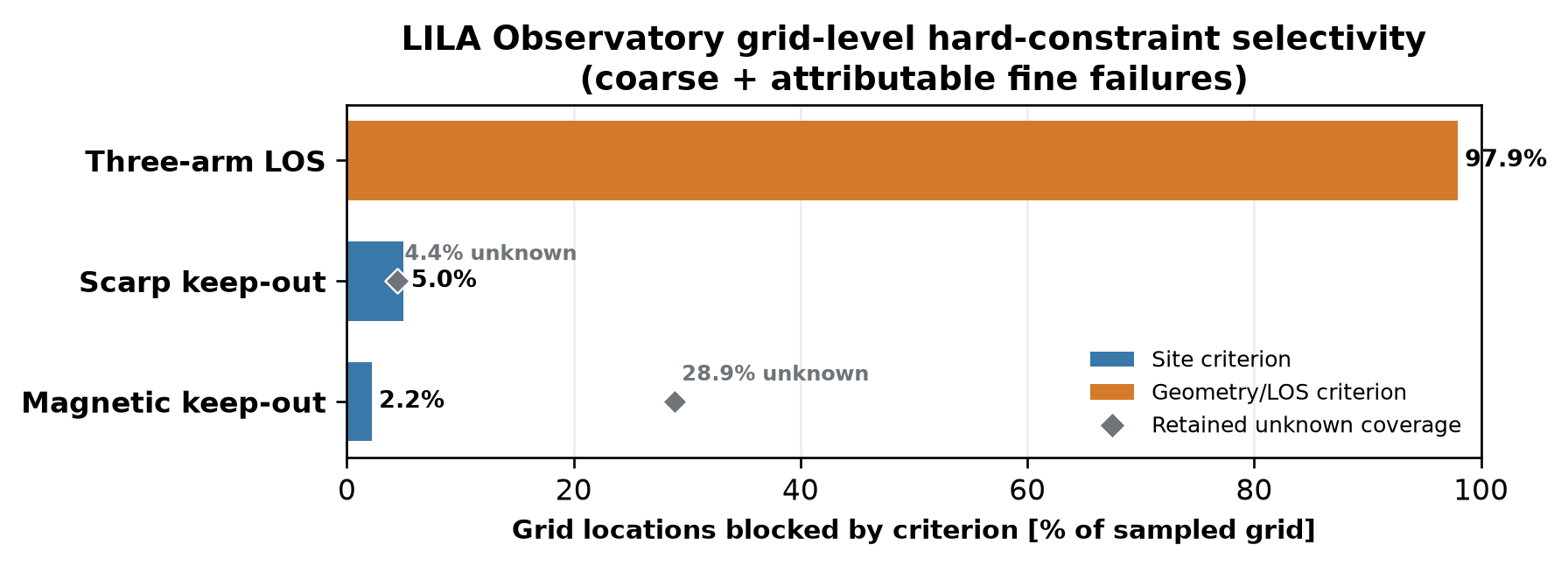}
}
\caption{\label{fig:Hard_Constraints}
  Grid-level hard-constraint selectivity for iLILA (top) and LILA Observatory (bottom). Orange bars denote geometry or LOS criteria and blue bars denote site criteria; grey diamonds show retained unknown coverage. Bars overlap because a grid seed may be blocked by more than one criterion. Observatory fine LOS failures are included, whereas unattributed iLILA regional failures remain in the attrition accounting.
  }
\end{figure}

Soft ranking influence is evaluated only within the final accepted grid catalog. One score term is removed at a time, configurations are re-ranked, and turnover in top-decile grid membership is recorded. For iLILA, removing LOS clearance replaces 68.2\% of the top decile and removing latitude replaces 47.7\%; roughness and scarp distance produce 13.6\% and 9.1\% turnover, while slope margin produces none. For the Observatory, latitude and scarp distance produce 87.5\% and 50.0\% turnover, respectively, while the small LOS-clearance term produces none in this catalog. Zero-weight regolith, magnetic-distance, and iLILA height-margin terms have zero influence by construction.

\begin{figure}
\centering
\resizebox{0.7\linewidth}{!}{
\includegraphics{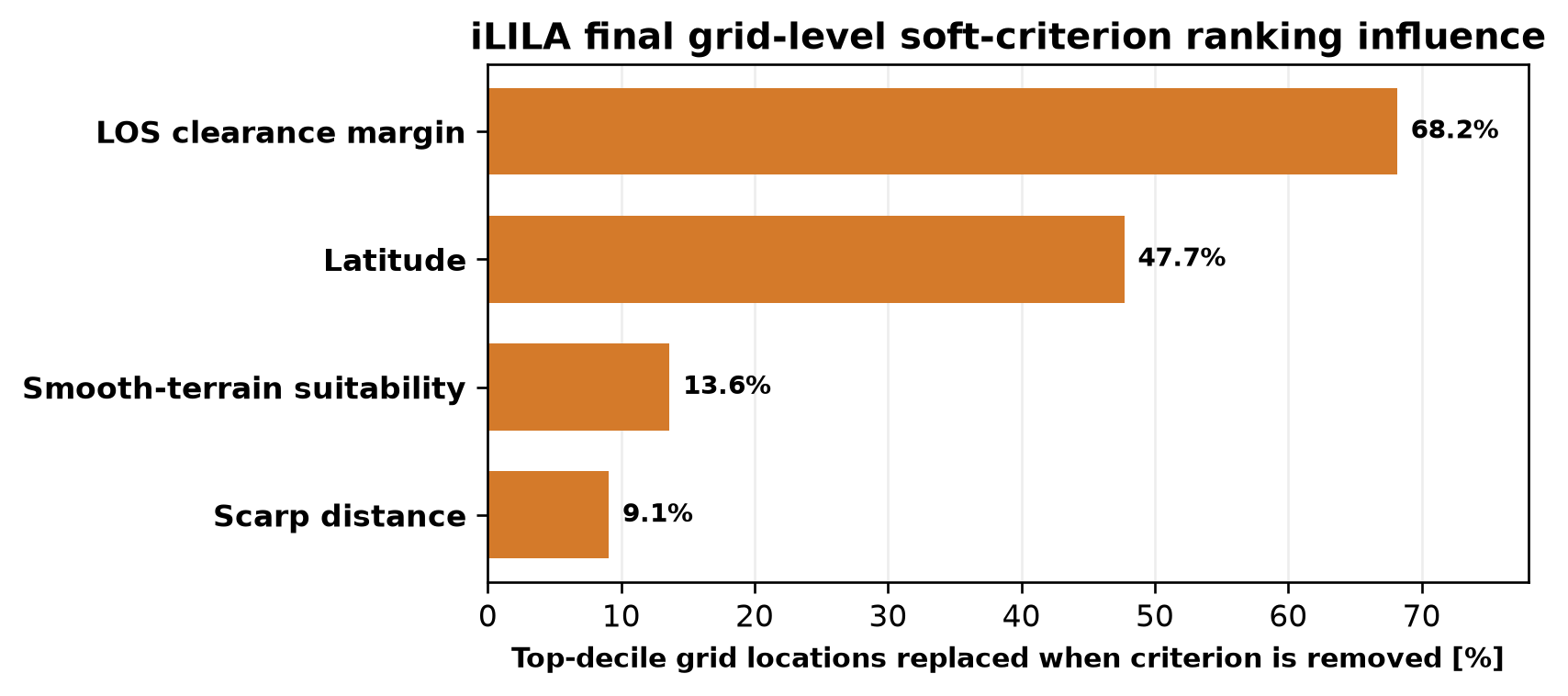}
}
\\[1ex]
\resizebox{0.7\linewidth}{!}{
\includegraphics{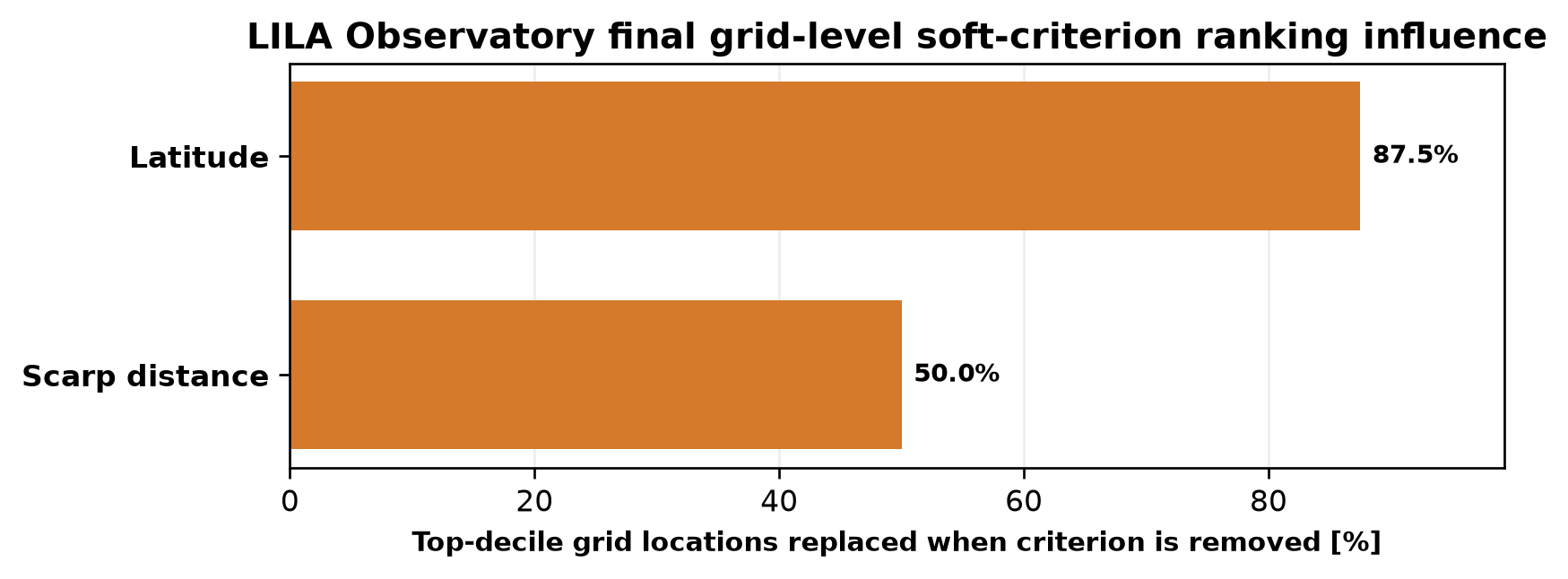}
}
\caption{\label{fig:Soft_Constraints}
  Top-decile grid turnover after removing each soft score term for final fine-accepted iLILA (top) and LILA Observatory (bottom) locations. This diagnostic measures practical ranking influence within the accepted catalog, not hard-constraint selectivity or causal importance.
  }
\end{figure}

Sensitivity studies were performed for both the hard constraints (Jaccard indices) and soft constraints (Spearman coefficients) to demonstrate stability of results to plausible changes in the various constraints. For $\pm20\%$ one-at-a-time hard-threshold perturbations, the minimum Jaccard similarity to the production catalog is 0.699 for iLILA—driven primarily by the endpoint height-difference requirement—and 0.959 for the LILA Observatory. By comparison, perturbing the magnetic-field cutoff and keep-out distance gives Jaccard similarities of 0.986–0.998 for iLILA and 0.972–1.000 for the Observatory, while changing the scarp keep-out distance gives 0.982–0.995 and 1.000, respectively; these environmental exclusions therefore have relatively little influence on the feasible-site population within the tested bounds. Soft selection sensitivity was studied by varying each coefficient from 0.5 to 1.5 times its production value. Spearman coefficient analysis showed that LOS margin ($\rho=0.94$) and latitude ($\rho=0.92$) dominated the overall reordering. Turnover at the top-decile was dominated by LOS margin, causing 38.6\% turnover compared to 18.2\% from latitude. Other constraints were much lower for both analyses. Overall, it is LOS and height difference requirements that drive sensitivity and not constraints that required significant  assumptions such as LTV requirements, keep-out radii, and electromagnetic anomaly strength.

\subsection{Dependence among criteria}

Dependence among hard-constraint failures is quantified using the phi coefficient, i.e. Pearson correlation applied to binary grid-level failure indicators. A value near $+1$ indicates coincident failures, 0 indicates little linear association, and $-1$ indicates anticorrelation. The calculation does not correlate individual configurations. In the all-seed iLILA panel, LTV route-slope and route-length failures are combined by logical OR as ``LTV suitability'' to avoid plotting their coincident grid outcomes twice. Most full-lattice associations are weak because the dominant criteria need not reject the same grid seed.

Fine-stage-only iLILA failures show stronger associations: endpoint height difference versus LOS has $\phi=0.64$, height difference versus LTV suitability has $\phi=0.95$, and LOS versus LTV suitability has $\phi=0.74$. These values are based on only 10, 6, and 11 failures, respectively, among 722 fine-inspected seeds and should not be interpreted as stable population relationships. For the Observatory fine stage, only the three-side LOS indicator varies (43 of 114 seeds fail), so no pairwise phi coefficient is defined.

\begin{figure}
\centering
\resizebox{0.49\linewidth}{!}{
\includegraphics{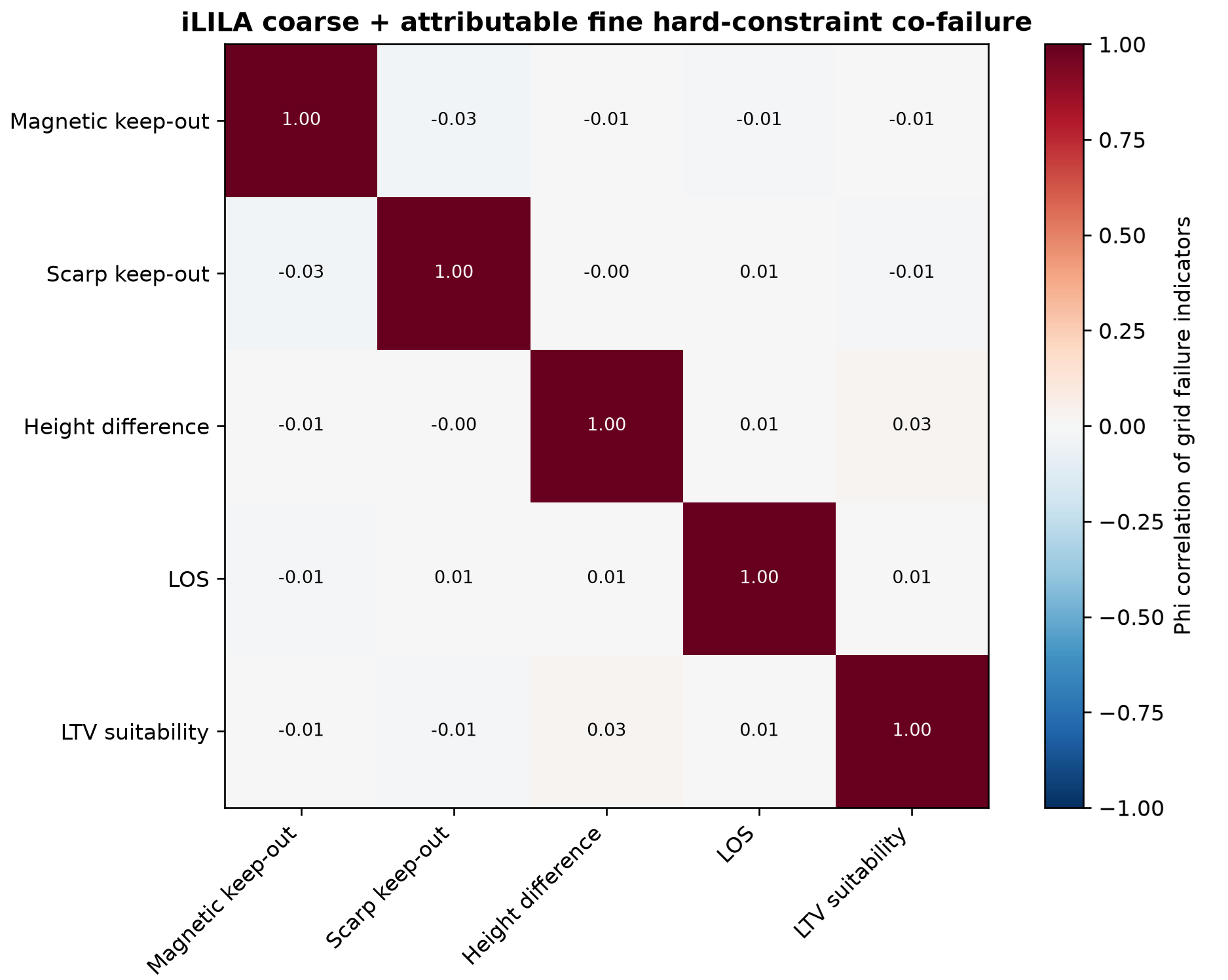}
}
\resizebox{0.49\linewidth}{!}{
\includegraphics{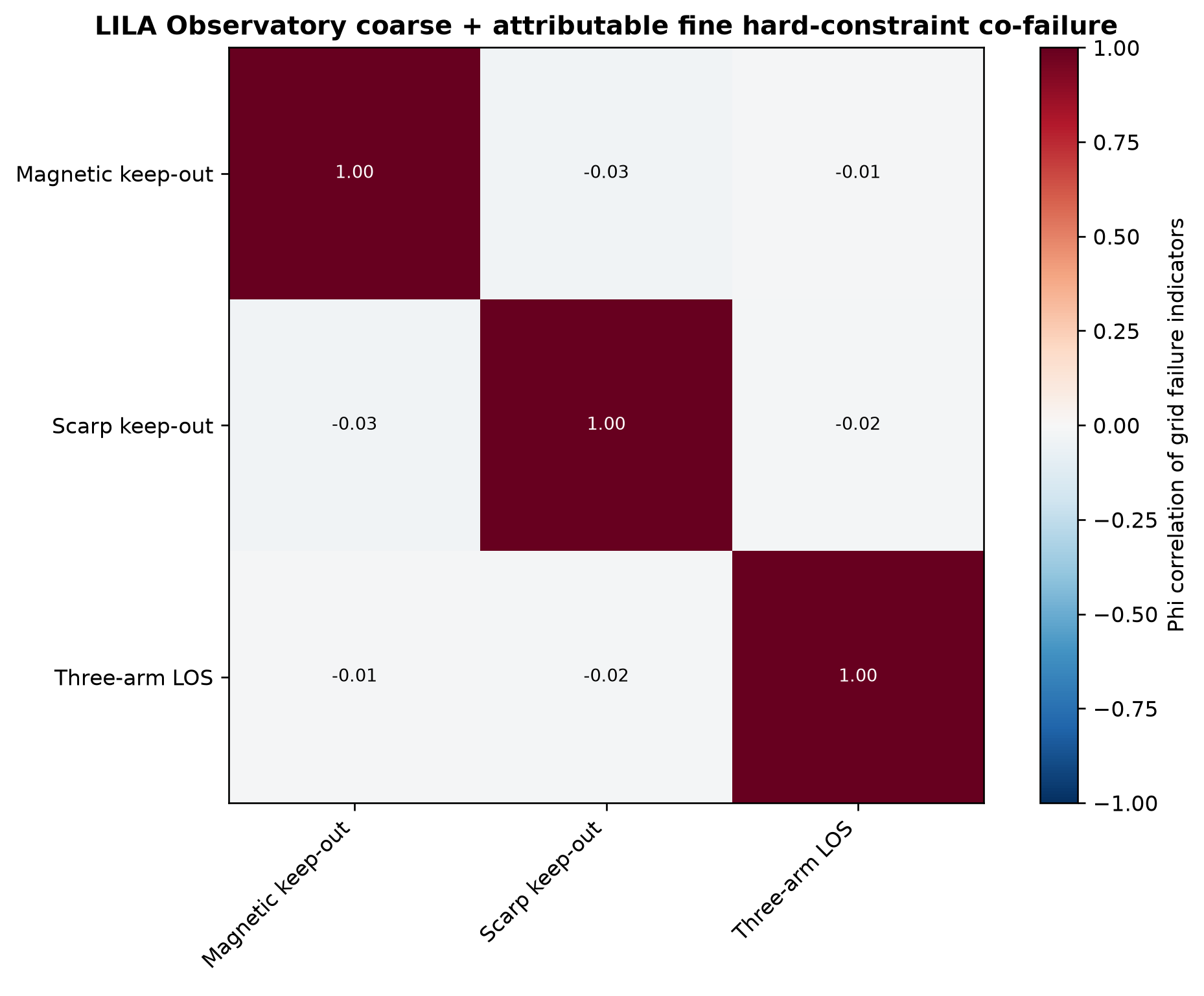}
}
\caption{\label{fig:Co-failure}
  Phi correlation of grid-level hard-failure indicators for iLILA (left) and LILA Observatory (right), using one record per original global seed. For iLILA, the physical indicators shown here are coarse-screen blockers; fine-stage associations are reported separately in the text. Observatory regional failures are included in its three-side LOS indicator.
  }
\end{figure}

The continuous analogue is the Pearson correlation between signed, weighted soft-score contributions for final accepted grid locations. Zero-weight terms are excluded. Positive coefficients indicate that two contributions tend to increase together; negative coefficients indicate a trade-off. For iLILA, the strongest association is between LOS-clearance contribution and smooth-terrain suitability ($r=-0.47$). This means that configurations with greater clearance tend to incur a larger roughness penalty. Observatory contribution correlations are weak; latitude and LOS-clearance contributions have $r=0.15$ in the inspected populations. These correlations describe the sampled catalog and do not establish causation.

\begin{figure}
\centering
\resizebox{0.49\linewidth}{!}{
\includegraphics{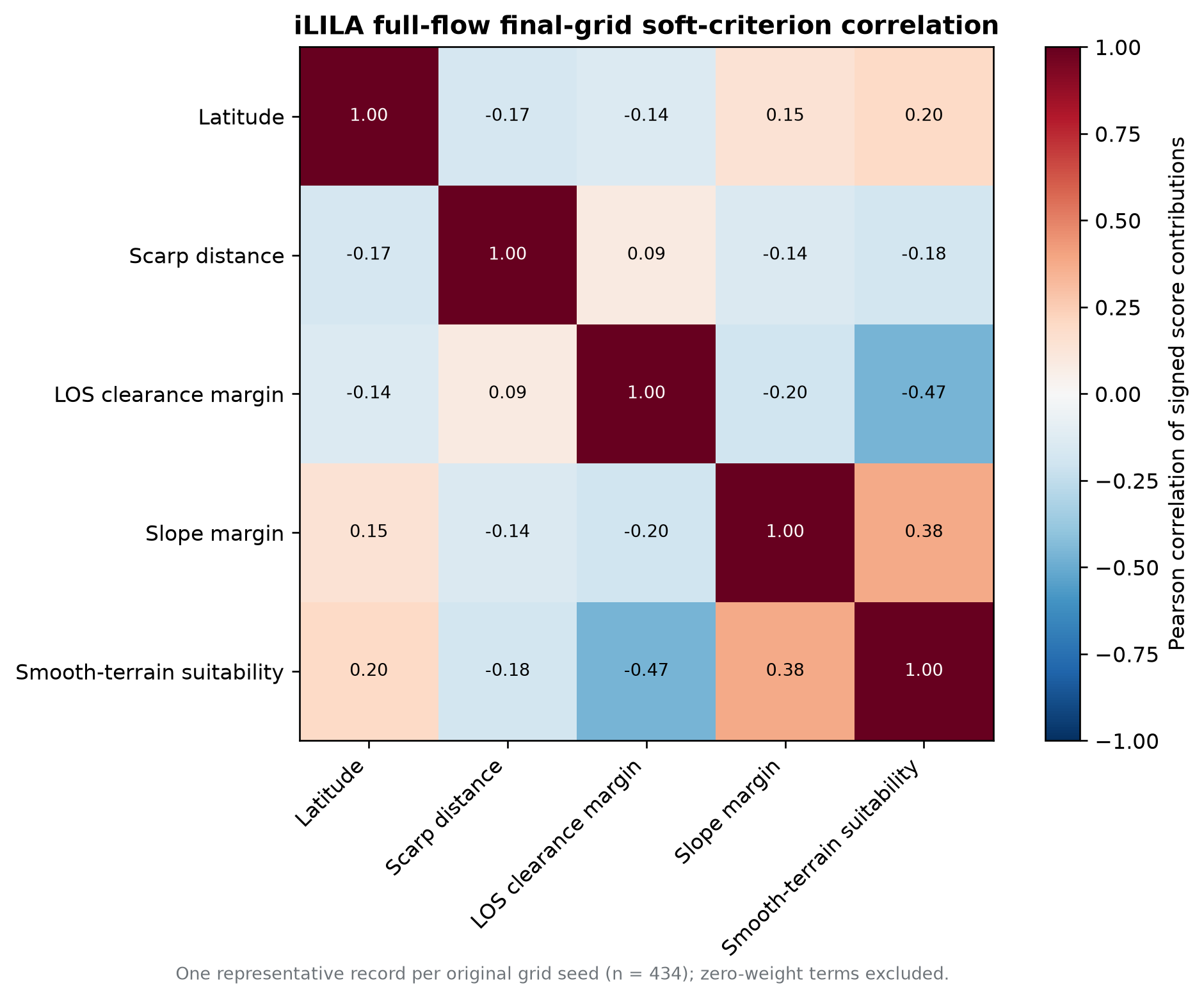}
}
\resizebox{0.49\linewidth}{!}{
\includegraphics{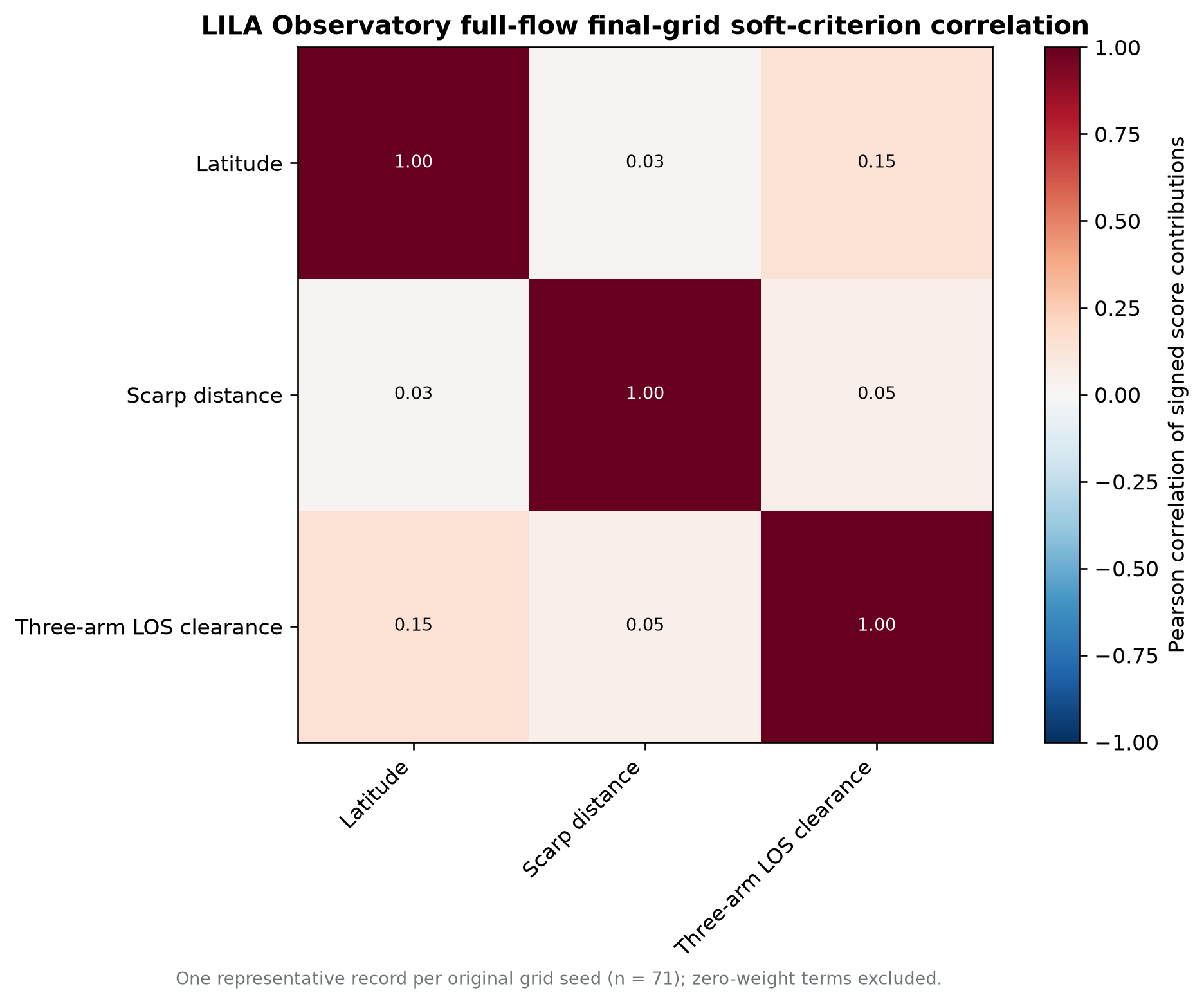}
}
\caption{\label{fig:Soft_correlation}
  Pearson correlation of signed active soft-score contributions for final fine-accepted iLILA (left) and LILA Observatory (right) grid locations. Each record is one original seed represented by its highest-ranked accepted configuration.
  }
\end{figure}

\section{Discussion and Conclusions}

The final catalogs contain 434 iLILA and 71 Observatory grid locations. iLILA therefore retains a comparatively broad candidate population under the registered assumptions, although exact placement is important: 288 of 722 coarse-accepted seeds fail the regional search. The spatial gaps over many mare plains and crater floors reflect the need for substantial endpoint relief, while the LTV constraints favor routes that remain smooth and traversable within that local relief. This combination, rather than generally rough terrain, governs compatibility with a co-manifested mission.

LILA Observatory is more selective but remains feasible in the modeled terrain: 71 circumcenters retain at least one regional triangle. The main challenge is simultaneous LOS along three sides, not proximity to a crater or the elevation of the computational center. The accepted configurations favor complex regional topography, but future landing precision, station-level slope, power, and deployment constraints could reduce or redistribute the catalog.

Four site-specific evidence needs remain before mission selection:
\begin{enumerate}
    \item a high-resolution stereo DTM and, ultimately, an in situ topographic survey to verify LOS, slope, roughness, and endpoint elevation;
    \item a credible estimate of depth to competent substrate for anchoring design;
    \item a site-specific model of shallow-moonquake and micrometeoroid-induced seismic noise in the measurement band; and
    \item a site-specific estimate of temperature-induced strain noise at the required sensitivity.
\end{enumerate}
The first is required to address sub-grid obstacles and provide a defensible clearance margin. The second would support a physically relevant anchoring criterion, unlike the present zero-weight surface-regolith descriptor. The last two would permit direct noise coupling estimates in place of the latitude and scarp proxies used here.

The search demonstrates feasibility for both concepts and provides a reproducible framework for comparing candidate locations as mission requirements evolve. Due to limitations in available data, final site selection must incorporate higher-resolution terrain, geotechnical information, mission architecture, and site-specific environmental noise. Still, credible layouts for both LDC- and GDC-style interferometers are readily found within the limitations of available data on the lunar surface. This avoids needing to construct scaffolding or tunnels to perform lunar GW interferometry. The framework can be further updated as new data becomes available and mission requirements evolve. 

\section*{Acknowledgments}

The authors acknowledge the use of OpenAI Codex (GPT-5.6 Sol) for code development, proofreading, figure preparation, and reference formatting.

P.L. acknowledges the support of Labex UnivEarth (ANR-10-LABX-0023 and ANR-18-IDEX-0001) and the Université Paris Cité Crossing Cutting Edges FIRE-UP project (ANR-21-EXES-0002). M.P.P. was supported by funds from the Jet Propulsion Laboratory, California Institute of Technology, under a contract with the National Aeronautics and Space Administration (80NM0018D0004). This work was supported by National Science Foundation grant NSF-2207999. K.J. and J.T. acknowledge
the support from the Vanderbilt University Office of Vice-Provost for Research and
Innovation.

\appendix
\section{Verification and validation}
\label{app:verification_validation}

The search was assessed using three distinct evidence classes: software verification, data validation, and comparison with external lunar data. Verification addressed whether the implemented mathematics and control flow reproduce the registered requirements at both coarse and fine stages. Validation was limited to the final fine-stage products and asks whether the inputs and accepted-site calculations are consistent with independent or higher-resolution evidence. These tests do not establish the physical accuracy of a candidate site.

\paragraph{External sets.}
In addition to the data sets in Section ~\ref{ss:lds}, external controls used published Apollo 11, 14 and 15 and Lunokhod~2 reflector coordinates and elevations, Kaguya LALT topography, a 2~m-posted LROC NAC Apollo~15 stereo DTM, and the Kaguya LMAG 30-km magnetic grid \cite{WilliamsFolkner2007_DE418,JAXA_KaguyaLALT_M,ArakiEtAl2009_KaguyaLALT,LROC_NAC_DTM_APOLLO15,HenriksenEtAl2017_LROCNACDTM,JAXA_KaguyaLMAG_30km,TsunakawaEtAl2014_LunarMagnetism}.

\paragraph{Verification and validation results.}
The frozen release passed all 32 verification checks and five final-stage threshold-conformance checks; its software test suite also passed 105 ordinary tests and 97 parameterized subtests without failure. Verification covered analytic geometry, coordinate and terrain calculations, LTV routing, thresholds and scoring, missing-data handling, coarse-to-fine traceability, robustness, deterministic recomputation and provenance; validation reopened and authenticated the declared regional DTMs and independently confirmed the applicable fine-stage metrics and thresholds for all 434 accepted iLILA records and 179 accepted Observatory triangles.

The separate external-evidence ledger contains 19 passes, two failures and five unresolved checks. Absolute coordinate/elevation extraction, large-margin Kaguya LOS classification, scale-matched NAC slope, local NAC roughness, magnetic classification, and available scarp-catalog controls passed. The two failures were the cross-method regolith-thickness transfer tests (2.244~m mean absolute error and Spearman $\rho=-0.097$); these led to setting the regolith weight coefficient is zero. Unresolved items are NAC LOS cases inside a validation-only $\pm8.2$~m ambiguity band around the 1~m model threshold, sub-grid slope hazards, depth to competent substrate, distance to complete polar scarp traces, and a survey-grade 1~m clearance guarantee. These data limitations primarily requires eventual survey-grade confirmation of the $1,\mathrm{m}$ limit as used in the paper and, if a polar site is eventually chosen, verification of polar scarp locations.

\paragraph{Impact on the reported conclusions.}
These results support the claims that the code implements the stated geometry and decision rules, that the frozen results are reproducible, and that both concepts have nonzero modeled feasibility on the tested lattice. They do not demonstrate construction readiness, feasible area, or meter-level physical clearance. The reported 1~m LOS criterion is a raster-model threshold. Promotion of any catalog entry to an engineering site requires a candidate-specific high-resolution stereo DTM or surface survey and confirmation of competent substrate for anchoring.

\section{Production data quality and use}
\label{app:data_quality}

\begin{table}[h]
\caption{Production data products and their exact use. ``Unknown allowed'' means that absence of mapped magnetic or scarp coverage does not reject a location; the location is flagged and receives the minimum corresponding preference score.}
\label{tab:production_data_quality}
\centering
{\scriptsize\sloppy
\setlength{\tabcolsep}{2pt}
\renewcommand{\arraystretch}{1.12}
\begin{tabular}{@{}p{0.12\textwidth}p{0.14\textwidth}p{0.13\textwidth}p{0.16\textwidth}p{0.23\textwidth}p{0.13\textwidth}@{}}
\hline
\textbf{Product} & \textbf{Version/identifier} & \textbf{Posting or effective resolution} & \textbf{Vertical/field uncertainty} & \textbf{Coverage and interpolation} & \textbf{Production role} \\
\hline
LOLA LDEM118 \cite{USGS_LOLA_LDEM118,NASAPDS_LOLA_GDR} & USGS 2014; PDS GDR v1.0; DOI 10.17189/\allowbreak1520642 & 256 pixels deg$^{-1}$; 118.45~m at equator; effective terrain resolution is track-density limited & 0.5~m per digital number is quantization only; no uniform per-pixel vertical error is published & Global interpolated grid from irregular LOLA tracks; equatorial track gaps commonly 1--2~km and locally about 4~km; queried bilinearly & Coarse elevation, endpoint relief, LOS, slope, roughness and LTV route tests \\
SLDEM2015 \cite{BarkerEtAl2016_SLDEM2015,MIT_SLDEM2015_Data} & 512-pixel-per-degree release; method DOI 10.1016/\allowbreak j.icarus.2015.07.039 & Approximately 59.2~m at equator & Typical vertical residual approximately 3--4~m after Terrain Camera registration to LOLA & $60^\circ$S--$60^\circ$N; SELENE Terrain Camera stereo merged and registered to LOLA; exact source pixels are stitched without resampling, then queried bilinearly & Nonpolar fine-stage iLILA terrain tests and Observatory three-side LOS \\
Adjusted polar LOLA DTMs \cite{NASAPGDA_LOLASouthPole,NASAPGDA_LOLAPoles} & PGDA products 90 and 98; DOIs 10.60903/\allowbreak gsfcpgda-lola-spole and 10.60903/\allowbreak gsfcpgda-lola-poles & Nominal 60~m posting; effective resolution varies with ground-track density & Spatially varying elevation-error products exist but are not propagated by the search; no single error applies & Source coverage from $60^\circ$ to each pole; bounded source-faithful windows; cross-processing of the same LOLA measurements used by the coarse DEM & Polar fine-stage iLILA terrain tests and Observatory three-side LOS \\
Lunar crustal magnetic-field map \cite{HoodEtAl2020_MagneticMapData} & PDS v1.0; DOI 10.17189/\allowbreak1520494 & $0.5^\circ$ latitude by $0.67^\circ$ longitude samples; smoothed effective resolution about $1.67^\circ\times3.33^\circ$ (roughly 50--100~km) & No single global nT uncertainty; zero-valued cells are treated as unreliable & $65^\circ$S--$65^\circ$N at 30~km altitude; nearest reliable cell must be within 100~km; missing coverage is allowed and flagged & Cells $\geq5$~nT define anomalies; 50~km hard keepout; ranking weight zero \\
LROC lobate-scarp polylines \cite{LROC_LobateScarps} & LROC RDR, 2014; no DOI assigned & Digitized as finely as 1:20,000; minimum mapped segments about 300~m; production traces densified to 1~km spacing & No formal positional-uncertainty raster & $60^\circ$S--$60^\circ$N; distance evaluated to densified mapped traces; missing coverage is allowed and flagged & 50~km hard keepout; normalized distance preference reaches full value at 200~km \\
LROC polar scarp locations \cite{LROC_PolarScarpLocations} & LROC RDR, 2015; no DOI assigned & Feature or cluster locations, not digitized fault traces & No formal positional uncertainty; distance-to-trace accuracy is unresolved & $65^\circ$--$90^\circ$ in both hemispheres; $60^\circ$--$65^\circ$ remains unmapped; missing coverage is allowed and flagged & Polar extension of the scarp keepout and distance preference; distance is a proxy \\
Regolith-thickness compilation \cite{RajsicEtAl2026_Regolith,JonesRajsic2026_RegolithData} & PSJ 2026; Zenodo v2.0.0, DOI 10.5281/\allowbreak zenodo.18829373 & 67 finite production points from 68 compiled estimates; not a raster resolution & Spatial leave-one-out MAE 1.872~m; cross-method MAE 2.244~m and rank test failed; not depth to competent substrate & Global spherical inverse-distance interpolation using nine nearest points and power two, with no maximum radius; nearest-measurement distance is retained & Reported anchoring proxy only; ranking weight zero and no hard rejection \\
\hline
\end{tabular}
}
\end{table}

Table~\ref{tab:production_data_quality} records the resolution, uncertainty, coverage and processing limitations that constrain interpretation of the search. Only products used by the production search are included; independent validation controls are described in appendix~\ref{app:verification_validation}.

\section*{References}

\bibliography{references}{}

@misc{NASAPDS_LOLA_GDR,
  author        = {{Neumann, G.A.}},
  title         = {{Lunar Orbiter Laser Altimeter Raw Data Set, LRO-L-LOLA-4-GDR-V1.0}},
  year          = {2010},
  doi           = {10.17189/1520642},
  howpublished  = {NASA Planetary Data System},
  url           = {https://pds.nasa.gov/ds-view/pds/viewDataset.jsp?dsid=LRO-L-LOLA-4-GDR-V1.0}
}

@misc{USGS_LOLA_LDEM118,
  author        = {{LOLA Science Team}},
  title         = {{Moon LRO LOLA DEM 118 m}},
  year          = {2014},
  url           = {https://astrogeology.usgs.gov/search/map/moon_lro_lola_dem_118m},
  howpublished  = {USGS Astrogeology Science Center}
}

@article{RajsicEtAl2026_Regolith,
  author        = {Raj\v{s}i\'c, Andrea and Daubar, Ingrid J. and Khan, Sierra and Jones,
                   Matthew J.},
  title         = {New Constraints on the Spatial and Temporal Evolution of the Lunar Surface
                   Regolith},
  journal       = {The Planetary Science Journal},
  year          = {2026},
  volume        = {7},
  number        = {7},
  pages         = {166},
  doi           = {10.3847/PSJ/ae7a6f}
}

@misc{JonesRajsic2026_RegolithData,
  author        = {Jones, Matthew J. and Raj\v{s}i\'c, Andrea},
  title         = {Lunar Surface Regolith Thickness Mapping},
  year          = {2026},
  version       = {2.0.0},
  doi           = {10.5281/zenodo.18829373},
  howpublished  = {Zenodo},
  url           = {https://zenodo.org/records/18829373}
}

@misc{HoodEtAl2020_MagneticMapData,
  author        = {Hood, Lon L. and Torres, Cesar B. and Oliveira, Joana S. and Wieczorek,
                   Mark A. and Stewart, Sarah T.},
  title         = {Lunar Crustal Magnetic Field Map},
  year          = {2020},
  doi           = {10.17189/1520494},
  howpublished  = {NASA Planetary Data System},
  url = {https://pds.nasa.gov/ds-view/pds/viewBundle.jsp?identifier=urn%3Anasa%3Apds%3Alunar-crust-magnetic.field-map&version=1.0}
}

@article{HoodEtAl2021_MagneticMap,
  author        = {Hood, Lon L. and Torres, Cesar B. and Oliveira, Joana S. and Wieczorek,
                   Mark A. and Stewart, Sarah T.},
  title         = {A New Large-Scale Map of the Lunar Crustal Magnetic Field and Its
                   Interpretation},
  journal       = {Journal of Geophysical Research: Planets},
  year          = {2021},
  volume        = {126},
  number        = {2},
  pages         = {1-17},
  doi           = {10.1029/2020JE006667},
}

@misc{LROC_LobateScarps,
  author        = {{LROC Science Operations Center}},
  title         = {Lunar Lobate Scarps (60 S to 60 N) Shapefile},
  year          = {2014},
  url           = {https://data.lroc.im-ldi.com/lroc/view_rdr/SHAPEFILE_LOBATE_SCARPS},
  howpublished  = {Lunar Reconnaissance Orbiter Camera Reduced Data Record}
}

@inproceedings{NelsonEtAl2014_LobateScarps,
  author        = {Nelson, D. M. and Koeber, S. D. and Daud, K. and Robinson, M. S. and
                   Watters, T. R. and Banks, M. E. and Williams, N. R.},
  title         = {Mapping Lunar Maria Extents and Lobate Scarps Using {LROC} Image Products},
  booktitle     = {45th Lunar and Planetary Science Conference},
  year          = {2014},
  pages         = {2861},
  url           = {https://www.hou.usra.edu/meetings/lpsc2014/pdf/2861.pdf},
}

@misc{LROC_PolarScarpLocations,
  author        = {{LROC Science Operations Center}},
  title         = {Locations of Polar Lunar Lobate Scarps},
  year          = {2015},
  url           = {https://data.lroc.im-ldi.com/lroc/view_rdr/SHAPEFILE_POLAR_SCARP_LOCATIONS},
  howpublished  = {Lunar Reconnaissance Orbiter Camera Reduced Data Record}
}

@article{WattersEtAl2015_LunarScarps,
  author        = {Watters, Thomas R. and Robinson, Mark S. and Collins, Geoffrey C. and
                   Banks, Maria E. and Daud, Katie and Williams, Nathan R. and Selvans,
                   Michelle M.},
  title         = {Global Thrust Faulting on the Moon and the Influence of Tidal Stresses},
  journal       = {Geology},
  year          = {2015},
  volume        = {43},
  number        = {10},
  pages         = {851--854},
  doi           = {10.1130/G37120.1},
}

@misc{NASAPGDA_LOLASouthPole,
  author        = {Barker, Michael K.},
  title         = {A New View of the Lunar South Pole from {LOLA}},
  year          = {2023},
  doi           = {10.60903/gsfcpgda-lola-spole},
  howpublished  = {NASA Goddard Planetary Geology, Geophysics and Geochemistry Laboratory}, 
  url = {https://pgda.gsfc.nasa.gov/products/90}
}

@article{BarkerEtAl2025_LOLAPolarRoughness,
  author        = {Barker, Michael K. and Mazarico, Erwan and Neumann, Gregory A. and Smith,
                   David E. and Zuber, Maria T. and Head, James W. and Sun, Xiaoli},
  title         = {Large-Scale Roughness Properties of the Lunar North and South Polar Regions
                   as Measured by the Lunar Orbiter Laser Altimeter ({LOLA})},
  journal       = {The Planetary Science Journal},
  year          = {2025},
  volume        = {6},
  pages         = {83},
  doi           = {10.3847/PSJ/adbc9d},
}

@misc{NASAPGDA_LOLAPoles,
  author        = {Barker, Michael K.},
  title         = {Large-Scale Roughness Properties of the Lunar Poles},
  year          = {2025},
  doi           = {10.60903/gsfcpgda-lola-poles},
  howpublished  = {NASA Goddard Planetary Geology, Geophysics and Geochemistry Laboratory},
  url = {https://pgda.gsfc.nasa.gov/products/98}
}

@article{BarkerEtAl2021_ImprovedLOLA,
  author        = {Barker, Michael K. and Mazarico, Erwan and Neumann, Gregory A. and Smith,
                   David E. and Zuber, Maria T. and Head, James W.},
  title         = {Improved {LOLA} Elevation Maps for South Pole Landing Sites: Error
                   Estimates and Their Impact on Illumination Conditions},
  journal       = {Planetary and Space Science},
  year          = {2021},
  volume        = {203},
  pages         = {105119},
  doi           = {10.1016/j.pss.2020.105119}
}

@article{BarkerEtAl2016_SLDEM2015,
  author        = {Barker, Michael K. and Mazarico, Erwan and Neumann, Gregory A. and Zuber,
                   Maria T. and Haruyama, Junichi and Smith, David E.},
  title         = {A New Lunar Digital Elevation Model from the Lunar Orbiter Laser Altimeter
                   and {SELENE} Terrain Camera},
  journal       = {Icarus},
  year          = {2016},
  volume        = {273},
  pages         = {346--355},
  doi           = {10.1016/j.icarus.2015.07.039}
}

@misc{MIT_SLDEM2015_Data,
  author        = {Barker, Michael K. and Mazarico, Erwan and Neumann, Gregory A. and Zuber,
                   Maria T. and Haruyama, Junichi and Smith, David E.},
  title         = {{SLDEM2015}: Lunar Digital Elevation Model},
  year          = {2015},
  url           = {https://imbrium.mit.edu/DATA/SLDEM2015/},
  howpublished  = {Massachusetts Institute of Technology LOLA Data Archive}
}

@techreport{WilliamsFolkner2007_DE418,
  author        = {Williams, James G. and Folkner, William M.},
  title         = {{DE418} Moon and Lunar Coordinates},
  year          = {2007},
  number        = {IOM 335-JGW,WMF-20071024-002},
  institution   = {Jet Propulsion Laboratory, California Institute of Technology},
  month         = {oct},
  url           = {https://naif.jpl.nasa.gov/pub/naif/generic_kernels/fk/satellites/de418_ephemeris_and_lunar_orientation_R1.pdf}
}

@misc{JAXA_KaguyaLALT_M,
  author        = {{JLALT Team}},
  title         = {{Kaguya/SELENE LALT Mare Serenitatis Regional Topography (LALT\_M)}},
  year          = {2007-2008},
  url           = {https://www.kaguya.jaxa.jp/en/science/LALT/The_lunar_topographic_data/LALT_M.zip},
  howpublished  = {Japan Aerospace Exploration Agency}}

@article{ArakiEtAl2009_KaguyaLALT,
  author        = {Araki, Hiroshi and others},
  title         = {Lunar Global Shape and Polar Topography Derived from {Kaguya-LALT} Laser
                   Altimetry},
  journal       = {Science},
  year          = {2009},
  volume        = {323},
  number        = {5916},
  pages         = {897--900},
  doi           = {10.1126/science.1164146},
}

@misc{LROC_NAC_DTM_APOLLO15,
  author        = {{LROC Science Operations Center}},
  title         = {Apollo 15 Landing Site Digital Terrain Model ({NAC\_DTM\_APOLLO15})},
  year          = {2020},
  version       = {1.9},
  url           = {https://data.lroc.im-ldi.com/lroc/view_rdr_product/NAC_DTM_APOLLO15},
  howpublished  = {NASA Planetary Data System, LRO-L-LROC-5-RDR-V1.0}
}

@article{HenriksenEtAl2017_LROCNACDTM,
  author        = {Henriksen, Mark R. and others},
  title         = {Extracting Accurate and Precise Topography from {LROC} Narrow Angle Camera
                   Stereo Observations},
  journal       = {Icarus},
  year          = {2017},
  volume        = {283},
  pages         = {122--137},
  doi           = {10.1016/j.icarus.2016.05.012},
}

@misc{JAXA_KaguyaLMAG_30km,
  author        = {{JAXA Kaguya LMAG Team}},
  title         = {{SELENE Moon LMAG Magnetic Anomaly Grid Option V1.0}},
  year          = {2014},
  url           = {https://data.darts.isas.jaxa.jp/pub/pds3/sln-l-lmag-5-ma-grid-option-v1.0/},
  howpublished  = {JAXA Data Archives and Transmission System}
}

@article{TsunakawaEtAl2014_LunarMagnetism,
  author        = {Tsunakawa, Hideo and Takahashi, Futoshi and Shimizu, Hisayoshi and Shibuya,s
                   Hidetoshi and Matsushima, Masaki},
  title         = {Regional Mapping of the Lunar Magnetic Anomalies at the Surface: Method and
                   Its Application to Strong and Weak Magnetic Anomaly Regions},
  journal       = {Icarus},
  year          = {2014},
  volume        = {228},
  pages         = {35--53},
  doi           = {10.1016/j.icarus.2013.09.026},
}

@article{Pou2024,
  author        = {Pou, Laurent and Panning, Mark P. and Styczinski, Marshall J. and Melwani
                   Daswani, Mohit and Nunn, Ceri and Vance, Steven D.},
  title         = {Tidal Seismicity in the Moon and Implications for the Rocky Interior of
                   Europa},
  journal       = {The Planetary Science Journal},
  year          = {2024},
  volume        = {5},
  number        = {6},
  pages         = {142},
  publisher     = {American Astronomical Society},
  month         = {jun},
  doi           = {10.3847/psj/ad47bc},
}

@article{Lognonn2019,
  author        = {Lognonn{\'e}, P. and others},
  title         = {SEIS: Insight's Seismic Experiment for Internal Structure of Mars},
  journal       = {Space Science Reviews},
  year          = {2019},
  volume        = {215},
  number        = {1},
  publisher     = {Springer Science and Business Media LLC},
  month         = {jan},
  doi           = {10.1007/s11214-018-0574-6},
  pages         = {1-170}
}

@article{Lognonn2020,
  author        = {Lognonn{\'e}, P. and others},
  title         = {Constraints on the shallow elastic and anelastic structure of Mars from
                   InSight seismic data},
  journal       = {Nature Geoscience},
  year          = {2020},
  volume        = {13},
  number        = {3},
  pages         = {213--220},
  publisher     = {Springer Science and Business Media LLC},
  month         = {feb},
  doi           = {10.1038/s41561-020-0536-y},
}

@article{Thor2020,
  author        = {Thor, Robin N. and Kallenbach, Reinald and Christensen, Ulrich R. and
                   Gl\"{a}ser, Philipp and Stark, Alexander and Steinbr\"{u}gge, Gregor and
                   Oberst, J\"{u}rgen},
  title         = {Determination of the lunar body tide from global laser altimetry data},
  journal       = {Journal of Geodesy},
  year          = {2020},
  volume        = {95},
  number        = {1},
  publisher     = {Springer Science and Business Media LLC},
  month         = {dec},
  doi           = {10.1007/s00190-020-01455-8},
  pages         = {1-14}
}

@article{Tsunakawa2010,
  author        = {Tsunakawa, Hideo and Shibuya, Hidetoshi and Takahashi, Futoshi and Shimizu,
                   Hisayoshi and Matsushima, Masaki and Matsuoka, Ayako and Nakazawa, Satoru
                   and Otake, Hisashi and Iijima, Yuichi},
  title         = {Lunar Magnetic Field Observation and Initial Global Mapping of Lunar
                   Magnetic Anomalies by MAP-LMAG Onboard SELENE (Kaguya)},
  journal       = {Space Science Reviews},
  year          = {2010},
  volume        = {154},
  pages         = {219--251},
  doi           = {10.1007/s11214-010-9652-0},
}

@article{LEFEUVRE20111,
  author        = {Mathieu {Le Feuvre} and Mark A. Wieczorek},
  title         = {Nonuniform cratering of the Moon and a revised crater chronology of the
                   inner Solar System},
  journal       = {Icarus},
  year          = {2011},
  volume        = {214},
  number        = {1},
  pages         = {1--20},
  doi           = {10.1016/j.icarus.2011.03.010},
}

@article{Lognonne2009,
  author        = {Lognonne, P. and Le Feuvre, M. and Johnson, C. L. and Weber, R. C.},
  title         = {Moon meteoritic seismic hum: Steady state prediction},
  journal       = {Journal of Geophysical Research},
  year          = {2009},
  volume        = {114},
  pages         = {E12003},
  doi           = {10.1029/2008JE003294},
}

@article{WATTERS2010,
  author        = {Thomas R. Watters and others},
  title         = {Evidence of Recent Thrust Faulting on the Moon Revealed by the Lunar
                   Reconnaissance Orbiter Camera},
  journal       = {Science},
  year          = {2010},
  volume        = {329},
  number        = {5994},
  pages         = {936--940},
  doi           = {10.1126/science.1189590},
}

@article{WILLIAMS2017300,
  author        = {J.-P. Williams and D.A. Paige and B.T. Greenhagen and E. Sefton-Nash},
  title         = {The global surface temperatures of the Moon as measured by the Diviner
                   Lunar Radiometer Experiment},
  journal       = {Icarus},
  year          = {2017},
  volume        = {283},
  pages         = {300--325},
  doi           = {10.1016/j.icarus.2016.08.012}
}

@article{Glanzer_2023,
  author        = {Glanzer, J and others},
  title         = {Data quality up to the third observing run of advanced LIGO: Gravity Spy
                   glitch classifications},
  journal       = {Classical and Quantum Gravity},
  year          = {2023},
  volume        = {40},
  number        = {6},
  pages         = {065004},
  publisher     = {IOP Publishing},
  month         = {feb},
  doi           = {10.1088/1361-6382/acb633},
}

@article{2020arXiv200708550J,
  author        = {Jani, Karan and Loeb, Abraham},
  title         = {Gravitational-Wave Lunar Observatory for Cosmology},
  journal       = {Journal of Cosmology and Astroparticle Physics},
  year          = {2021},
  volume        = {2021},
  number        = {06},
  pages         = {044},
  doi           = {10.1088/1475-7516/2021/06/044},
}

@article{Horanyi:2015,
  author        = {{Hor{\'a}nyi}, M. and {Szalay}, J.~R. and {Kempf}, S. and {Schmidt}, J. and
                   {Gr{\"u}n}, E. and {Srama}, R. and {Sternovsky}, Z.},
  title         = {{A permanent, asymmetric dust cloud around the Moon}},
  journal       = {Nature},
  year          = {2015},
  volume        = {522},
  number        = {7556},
  pages         = {324--326},
  month         = {jun},
  doi           = {10.1038/nature14479},
}

@article{Harms:2022,
  author        = {{Harms}, Jan},
  title         = {{Seismic Background Limitation of Lunar Gravitational-Wave Detectors}},
  journal       = {Phys. Rev. Lett.},
  year          = {2022},
  volume        = {129},
  number        = {7},
  pages         = {071102},
  month         = {aug},
  doi           = {10.1103/PhysRevLett.129.071102},
  eid           = {071102},
}

@article{Harms:2021,
  author        = {{Harms}, Jan and others},
  title         = {{Lunar Gravitational-wave Antenna}},
  journal       = {Astrophysical Journal},
  year          = {2021},
  volume        = {910},
  number        = {1},
  pages         = {1},
  month         = {mar},
  doi           = {10.3847/1538-4357/abe5a7},
  eid           = {1},
}

@article{panning2025potentiallunarinteriorscience,
  author        = {Panning, Mark P. and Lognonn{\'e}, Philippe and Creighton, Teviet and
                   Trippe, James and Quetschke, Volker and Majstorovi{\'c}, Josipa and Jani,
                   Karan},
  title         = {Potential for Lunar Interior Science by the Gravitational-Wave Detector
                   {LILA}},
  journal       = {Classical and Quantum Gravity},
  year          = {2026},
  volume        = {43},
  pages         = {127001},
  doi           = {10.1088/1361-6382/ae67ca},
}

@misc{shapiro2025vibrationisolationlaserinterferometer,
  author        = {Shapiro, Brett N.},
  title         = {Vibration Isolation for the Laser Interferometer Lunar Antenna},
  year          = {2025},
  eprint        = {2512.11268},
  archiveprefix = {arXiv},
  primaryclass  = {gr-qc},
}

@article{Rubanu_2009,
  author        = {Rubanu, F and Poggiani, R and Hough, J},
  title         = {Interplanetary dust: a source of noise for LISA?},
  journal       = {Classical and Quantum Gravity},
  year          = {2009},
  volume        = {26},
  number        = {22},
  pages         = {225012},
  month         = {oct},
  doi           = {10.1088/0264-9381/26/22/225012},
}

@techreport{NASA2023,
  author        = {Phillip B. Abel and others},
  title         = {Lunar Dust Mitigation: A Guide and Reference: First Edition (2021)},
  year          = {2023},
  number        = {TP-20220018746},
  institution   = {National Aeronautics and Space Administration},
  address       = {Cleveland, Ohio and Kennedy Space Center, Florida and Houston, Texas and
                   Hampton, Virginia},
  month         = {sep},
  url           = {https://ntrs.nasa.gov/api/citations/20220018746/downloads/TP-20220018746.pdf},
}

@article{https://doi.org/10.1002/2013JE004559,
  author        = {Williams, James G. and others},
  title         = {Lunar interior properties from the GRAIL mission},
  journal       = {Journal of Geophysical Research: Planets},
  year          = {2014},
  volume        = {119},
  number        = {7},
  pages         = {1546--1578},
  doi           = {10.1002/2013JE004559},
}

@article{https://doi.org/10.1029/2001JE001506,
  author        = {Feldman, W. C. and Gasnault, O. and Maurice, S. and Lawrence, D. J. and
                   Elphic, R. C. and Lucey, P. G. and Binder, A. B.},
  title         = {Global distribution of lunar composition: New results from Lunar Prospector},
  journal       = {Journal of Geophysical Research: Planets},
  year          = {2002},
  volume        = {107},
  number        = {E3},
  pages         = {5--1--5--14},
  doi           = {10.1029/2001JE001506},
}

@techreport{apl_lsii_2025_lit,
  author        = {{APL LSII Systems Integrator Team}},
  title         = {Lunar Infrastructure Technologies ({LIT}) Overview},
  year          = {2025},
  institution   = {Johns Hopkins University Applied Physics Laboratory},
  type          = {Technical Report},
  month         = {October},
  url           = {https://lsic.jhuapl.edu/uploadedDocs/documents/2891-LIT_Overview_20250924.pdf},
}

@article{ARAUJO2005451,
  author        = {H.M. Ara{\'u}jo and P. Wass and D. Shaul and G. Rochester and T.J. Sumner},
  title         = {Detailed calculation of test-mass charging in the LISA mission},
  journal       = {Astroparticle Physics},
  year          = {2005},
  volume        = {22},
  number        = {5},
  pages         = {451--469},
  doi           = {10.1016/j.astropartphys.2004.09.004},
}

@misc{LISAmain,
  author        = {Amaro-Seoane, Pau and others},
  title         = {Laser Interferometer Space Antenna},
  year          = {2017},
  eprint        = {1702.00786},
  archiveprefix = {arXiv},
  primaryclass  = {astro-ph.IM}
}

@techreport{LIGO-Instrument2018,
  author        = {{LIGO Scientific Collaboration}},
  title         = {{LSC} Instrument Science White Paper 2018},
  year          = {2018},
  number        = {LIGO-T1800133},
  institution   = {LIGO Scientific Collaboration},
  url           = {https://dcc.ligo.org/public/0151/T1800133/004/T1800133-instrument-science-white-v4.pdf},
}

@techreport{NASA_SP_20205009602_2020,
  author        = {{National Aeronautics and Space Administration}},
  title         = {Artemis {III} Science Definition Team Report},
  year          = {2020},
  number        = {NASA/SP-20205009602},
  institution   = {National Aeronautics and Space Administration},
  month         = {dec},
  url           = {https://www.nasa.gov/wp-content/uploads/2015/01/artemis-iii-science-definition-report-12042020c.pdf},
}

@article{SATO2017216,
  author        = {Hiroyuki Sato and Mark S. Robinson and Samuel J. Lawrence and Brett W.
                   Denevi and Bruce Hapke and Bradley L. Jolliff and Harald Hiesinger},
  title         = {Lunar mare TiO2 abundances estimated from UV/Vis reflectance},
  journal       = {Icarus},
  year          = {2017},
  volume        = {296},
  pages         = {216--238},
  doi           = {10.1016/j.icarus.2017.06.013},
}

@article{Milliken2017,
  author        = {Milliken, Ralph E. and Li, Shuai},
  title         = {Remote detection of widespread indigenous water in lunar pyroclastic
                   deposits},
  journal       = {Nature Geoscience},
  year          = {2017},
  volume        = {10},
  number        = {8},
  pages         = {561--565},
  month         = {aug},
  doi           = {10.1038/ngeo2993},
}

@article{Horvath2022,
  author        = {Horvath, Tyler and Hayne, Paul O. and Paige, David A.},
  title         = {Thermal and Illumination Environments of Lunar Pits and Caves: Models and
                   Observations From the Diviner Lunar Radiometer Experiment},
  journal       = {Geophysical Research Letters},
  year          = {2022},
  volume        = {49},
  number        = {14},
  pages         = {e2022GL099710},
  doi           = {10.1029/2022GL099710}
}

@misc{NASA_ltv,
  author        = {{National Aeronautics and Space Administration}},
  title         = {Lunar Terrain Vehicle},
  year          = {2026},
  url           = {https://www.nasa.gov/moonbase/lunar-terrain-vehicle/},
  howpublished  = {NASA Moon Base},
}

@misc{LIGO:2012aa,
  author        = {{LIGO Scientific Collaboration} and {Virgo Collaboration}},
  title         = {Sensitivity Achieved by the {LIGO} and Virgo Gravitational-Wave Detectors
                   during {LIGO}'s Sixth and Virgo's Second and Third Science Runs},
  year          = {2012},
  eprint        = {1203.2674},
  archiveprefix = {arXiv},
  primaryclass  = {gr-qc},
}

@article{DEANGELIS2007169,
  author        = {G. {De Angelis} and F.F. Badavi and J.M. Clem and S.R. Blattnig and M.S.
                   Clowdsley and J.E. Nealy and R.K. Tripathi and J.W. Wilson},
  title         = {Modeling of the Lunar Radiation Environment},
  journal       = {Nuclear Physics B - Proceedings Supplements},
  year          = {2007},
  volume        = {166},
  pages         = {169--183},
  doi           = {10.1016/j.nuclphysbps.2006.12.034}
}

@techreport{EEE-INST-002,
  author        = {Sahu, Kusum},
  title         = {{EEE-INST-002}: Instructions for {EEE} Parts Selection, Screening,
                   Qualification, and Derating},
  year          = {2008},
  number        = {NASA/TP-2003-212242},
  institution   = {NASA Goddard Space Flight Center},
  address       = {Greenbelt, MD},
  month         = {apr},
  url           = {https://nepp.nasa.gov/pages/EEE-INST-002.cfm}
}

@techreport{MIL-PRF-38534,
  author        = {{United States Department of Defense}},
  title         = {Hybrid Microcircuits, General Specification For},
  year          = {2024},
  number        = {MIL-PRF-38534},
  institution   = {Defense Logistics Agency Land and Maritime},
  address       = {Columbus, OH},
  month         = {nov},
  url           = {https://quicksearch.dla.mil/qsDocDetails.aspx?ident_number=67840}
}

@article{https://doi.org/10.1029/2020JE006623,
  author        = {Feng, Jianqing and Siegler, Matthew A.},
  title         = {Reconciling the Infrared and Microwave Observations of the Lunar South
                   Pole: A Study on Subsurface Temperature and Regolith Density},
  journal       = {Journal of Geophysical Research: Planets},
  year          = {2021},
  volume        = {126},
  number        = {9},
  pages         = {1-14},
  doi           = {10.1029/2020JE006623}
}

@techreport{NASA_PTC_Guidebook_4_0_2023,
  author        = {Liles, Kaitlin and Amundsen, Ruth},
  title         = {{NASA} Passive Thermal Control Engineering Guidebook},
  year          = {2023},
  number        = {20230013900},
  institution   = {National Aeronautics and Space Administration},
  version       = {4.0},
  url           = {https://ntrs.nasa.gov/citations/20230013900},
}

@article{MAZARICO20111066,
  author        = {E. Mazarico and G.A. Neumann and D.E. Smith and M.T. Zuber and M.H.
                   Torrence},
  title         = {Illumination conditions of the lunar polar regions using LOLA topography},
  journal       = {Icarus},
  year          = {2011},
  volume        = {211},
  number        = {2},
  pages         = {1066--1081},
  doi           = {10.1016/j.icarus.2010.10.030},
}

@article{Aasi2015AdvancedLIGO,
  author        = {Aasi, J. and others},
  collaboration = {LIGO Scientific Collaboration},
  title         = {The Advanced LIGO Detectors},
  journal       = {Classical and Quantum Gravity},
  year          = {2015},
  volume        = {32},
  number        = {7},
  pages         = {074001},
  doi           = {10.1088/0264-9381/32/7/074001},
}

@article{Williams2019,
  author        = {Williams, J.-P. and Greenhagen, B. T. and Paige, D. A. and Schorghofer, N.
                   and Sefton-Nash, E. and Hayne, P. O. and Lucey, P. G. and Siegler, M. A.
                   and Aye, K. Michael},
  title         = {Seasonal Polar Temperatures on the Moon},
  journal       = {Journal of Geophysical Research: Planets},
  year          = {2019},
  volume        = {124},
  number        = {10},
  pages         = {2505--2521},
  doi           = {10.1029/2019JE006028},
}

@article{REITZ201278,
  author        = {Guenther Reitz and Thomas Berger and Daniel Matthiae},
  title         = {Radiation exposure in the moon environment},
  journal       = {Planetary and Space Science},
  year          = {2012},
  volume        = {74},
  number        = {1},
  pages         = {78--83},
  doi           = {10.1016/j.pss.2012.07.014}
}

@article{GLASER201478,
  author        = {P. Gl{\"a}ser and F. Scholten and D. {De Rosa} and R. {Marco Figuera} and
                   J. Oberst and E. Mazarico and G.A. Neumann and M.S. Robinson},
  title         = {Illumination conditions at the lunar south pole using high resolution
                   Digital Terrain Models from LOLA},
  journal       = {Icarus},
  year          = {2014},
  volume        = {243},
  pages         = {78--90},
  doi           = {10.1016/j.icarus.2014.08.013},
}

@article{Johnston2003,
  author        = {Johnston, A.H.},
  title         = {Radiation effects in light-emitting and laser diodes},
  journal       = {IEEE Transactions on Nuclear Science},
  year          = {2003},
  volume        = {50},
  number        = {3},
  pages         = {689--703},
  doi           = {10.1109/TNS.2003.812926},
}

@techreport{Katzan1991,
  author        = {Katzan, Cynthia M. and Edwards, Jonathan L.},
  title         = {Lunar Dust Transport and Potential Interactions with Power System
                   Components},
  year          = {1991},
  number        = {NASA-CR-4404},
  institution   = {National Aeronautics and Space Administration},
  type          = {Contractor Report},
  address       = {Washington, DC},
  month         = {nov},
  doi           = {10.2172/10181067},
}

@misc{Creighton2025,
  author        = {Creighton, Teviet and Lognonn{\'e}, Philippe and Panning, Mark P. and
                   Trippe, James and Quetschke, Volker and Jani, Karan},
  title         = {Fundamental Noise and Gravitational-Wave Sensitivity of the Laser
                   Interferometer Lunar Antenna ({LILA})},
  year          = {2025},
  eprint        = {2508.18437},
  archiveprefix = {arXiv},
  primaryclass  = {gr-qc},
}

@incollection{LOGNONNE201565,
  author        = {P. Lognonn{\'e} and C.L. Johnson},
  editor        = {Gerald Schubert},
  title         = {10.03 - Planetary Seismology},
  booktitle     = {Treatise on Geophysics (Second Edition)},
  year          = {2015},
  pages         = {65--120},
  edition       = {Second Edition},
  publisher     = {Elsevier},
  address       = {Oxford},
  doi           = {10.1016/B978-0-444-53802-4.00167-6},
  isbn          = {978-0-444-53803-1},
}

@article{Lemoine2014GRGM900C,
  author        = {Lemoine, Frank G. and others},
  title         = {{GRGM900C}: A degree 900 lunar gravity model from {GRAIL} primary and
                   extended mission data},
  journal       = {Geophysical Research Letters},
  year          = {2014},
  volume        = {41},
  pages         = {3382-3389},
  doi           = {10.1002/2014GL060027},
}

@article{Wieczorek2013GRAILCrust,
  author        = {Wieczorek, Mark A. and others},
  title         = {The crust of the Moon as seen by {GRAIL}},
  journal       = {Science},
  year          = {2013},
  volume        = {339},
  number        = {6120},
  pages         = {671--675},
  doi           = {10.1126/science.1231530},
}

@misc{GRAILDerivedGravityPDS,
  author        = {D. S. Kahan},
  title         = {{GRAIL} Moon {LGRS} Derived Gravity Science Data Products V1.0, {GRAIL-L-LGRS-5-RDR-V1.0}},
  year          = {2013},
  publisher     = {NASA Planetary Data System},
  doi           = {10.17189/1519529}, 
  url           = {https://pds.nasa.gov/ds-view/pds/viewDataset.jsp?dsid=GRAIL-L-LGRS-5-RDR-V1.0}
}

@article{Horanyi2014,
  author        = {Hor{\'a}nyi, M. and others},
  title         = {The Lunar Dust Experiment (LDEX) Onboard the Lunar Atmosphere and Dust
                   Environment Explorer (LADEE) Mission},
  journal       = {Space Science Reviews},
  year          = {2014},
  volume        = {185},
  number        = {1},
  pages         = {93--113},
  doi           = {10.1007/s11214-014-0118-7},
}

\end{document}